 \documentclass[amsmath,amssymb,floatfix]{revtex4}

\usepackage{graphicx}
\usepackage{dcolumn} 
\usepackage{bm}      

\usepackage{color}

\begin{document}
\title{The interplay between helicity and rotation in turbulence: \ \\
\      implications for scaling laws and small-scale dynamics}
\markboth{The interplay between helicity and rotation in turbulent flows}
         {The interplay between helicity and rotation in turbulent flows}
\author{A. Pouquet$^{1,2}$ and P.D. Mininni$^{1,3}$}
\affiliation{$^1$Computational and Information Systems Laboratory, NCAR, 
         P.O. Box 3000, Boulder, Colorado 80307-3000, U.S.A. \\
             $^2$Earth and Sun Systems Laboratory, NCAR, P.O. Box 3000, 
         Boulder, Colorado 80307-3000, U.S.A. \\
             $^3$ Departamento de F\'\i sica, Facultad de Ciencias Exactas 
         y Naturales, Universidad de Buenos Aires and CONICET, Ciudad 
         Universitaria, 1428 Buenos Aires, Argentina. }       
\date{\today}

\begin{abstract}
Invariance properties of physical systems govern their behavior: energy conservation in turbulence drives a wide distribution of energy among modes, observed in geophysical or astrophysical flows. In ideal hydrodynamics, the role of helicity conservation (correlation between velocity and its curl, measuring departures from mirror symmetry) remains unclear since it does not alter the energy spectrum. However, with solid body rotation, significant differences emerge between helical and non-helical flows. We first outline several results, like the energy and helicity spectral distribution and the breaking of strict universality for the individual spectra. Using massive numerical simulations, we then show that small-scale structures and their intermittency properties differ according to whether helicity is present or not, in particular with respect to the emergence of Beltrami-core vortices (BCV) that are laminar helical vertical updrafts. These results point to the discovery of a small parameter besides the Rossby number; this could relate the problem of rotating helical turbulence to that of critical phenomena, through renormalization group and weak turbulence theory. This parameter can be associated with the adimensionalized ratio of the energy to helicity flux to small scales, the three-dimensional energy cascade being weak and self-similar.
\end{abstract}

\maketitle

\section{Introduction}  \label{s:intro}

Turbulent flows, ubiquitous in nature, defy analysis due to their inherent complexity because of their nonlinearities leading to the strong coupling of a very large number of modes. Scaling laws beyond second-order, the order dealing with energy distribution in an incompressible fluid, are not amenable to simple dimensional analysis unless the flow is self-similar. Self-similar behavior in turbulent flows has been observed until now only at large scale, in the so-called inverse cascades where excitations injected at an intermediate wavenumber reach lower modes as time elapses; the inverse cascade of energy in two-dimensional Navier-Stokes turbulence (see e.g., Kraichnan and Montgomery 1980; McWilliams 1984; Smith and Waleffe 1999;  Boffetta et al. 2000; Tabeling 2002; Boffeta 2007), or the inverse cascade of magnetic potential in two dimensional and of magnetic helicity in three-dimensional magnetohydrodynamics 
are but three examples of this phenomenon. The link between inverse cascade, which emphasizes the emergence of order out of an otherwise chaotic flow as far as small-scale properties are concerned, and singularity of the underlying primitive equations is not clear. For two-dimensional Navier-Stokes turbulence (2D-NS hereafter)  the regularity for all times can be attributed to the conservation of enstrophy, or squared vorticity $\left<|\mbox{\boldmath $\omega$}|^2\right>$ with $\mbox{\boldmath $\omega$}=\nabla \times {\bf u}$ the vorticity and ${\bf u}$ the velocity field; but regularity in the case of magnetohydrodynamics (MHD), in two or three dimensions, is an open problem. For 2D-NS as well as for surface quasi-geostrophic turbulence for a rotating stably stratified layer  (Pierrehumbert et al. 1994), the self-similar behavior in the formation of eddies at scales larger than the energy input scale has been recently attributed, when examining the scaling properties of iso-vorticity lines, to conformal invariance, i.e. local scale invariance through transformations that preserve angles but not distances (Bernard et al. 2006, 2007). In that case, a link found to percolation theory allows for the analytical determination of scaling exponents such as the fractal dimension of vortex clusters, although the consequences for classical statistical measures of turbulent flows, e.g., through scaling laws for correlation and structure functions, has not been clarified yet. However, the three-dimensional case is known to be much more complex. In three dimensional homogeneous and isotropic fluid turbulence the flow is not scale invariant and the knowledge of one scaling exponent does not allow for the prediction of the exponents for all orders. The search for self-similar quantities in three-dimensional turbulence is a long-standing problem; it would relate its study with critical phenomena and the out-of-equilibrium statistics of systems with a large number of modes, and it would allow the use of tools -- such as the renormalization group (Ma and Mazenko 1975) -- from quantum field theory, condensed matter and statistical mechanics to further our understanding of such flows. 

It is well-known that, under the influence of a strong external agent such as gravity, rotation or magnetic fields, several features emerge: on the one hand, waves are present in the flow (gravity, inertial, or Alfv\'en waves respectively) and the interactions between nonlinear eddies and waves are not fully understood. On the other hand, the flow becomes anisotropic and in fact tends to (although never reaches) a near two-dimensional state, with thin layers in the case of stratification, or columnar (Taylor) vortices in the case of rotation. Since two-dimensional and three-dimensional turbulent flows have vastly different dynamics, one may ask in what way are the three-dimensional dynamics of a flow under such circumstances altered, and in what way they may possibly be linked to the much simpler two-dimensional case, simpler be it only because regularity for all time is proven as mentioned before.

In this context, we propose in this paper a brief review of rotating turbulent flows and specifically of the role that helicity (the correlation between velocity and vorticity) may play in such a dynamics. Rotating flows have been studied in the laboratory and observed in atmospheric flows as well as planets and stars (Greenspan 1968; Pedlosky 1986); a recent review of experiments performed on rotating turbulence, from 1975 to the present and both in the decay and in the statistically steady cases, can be found in van Bokhoven et al. (2009) (see Table 1 of their paper). Rotating flows have been investigated as well with theoretical tools and models of turbulence such as Large Eddy Simulations (Sagaut and Cambon 2008), and with direct numerical simulations (for recent studies of the specific effects of rotation on shear flows or on convective flows, see respectively Jacobitz et al. 2008 and Zhong et al. 2009). For example, in the atmosphere, close-range observations (2 km) with large resolution (50 m) during the VORTEX campaign (Verification of the Origin of Rotation in Tornadoes EXperiments; Markovski et al. 1998) using Doppler On Wheels radars with deployed range of at most 120 s revealed many fine-scale structures, observed under the name of multi-vortex cores (Wurman et al. 1996): a quiet (laminar) eye, 200 m wide, surrounded by a ring of debris (implying turbulence) of similar width; moreover, at times, several concentric structures were observed as well, interpreted as different classes of debris with differing radar reflectivities. It has been known for a long time that multiple vortices are present in a tornado, for example from examination after the fact of the damages caused by the passage of the tornado; it is only recently that observational techniques have been able to quantify dynamical variables and localize such small-scale intense structures. In a recent study (Wurman 2002) it was shown for example that such multiple vortices are persistent and surround the central eye, with shear across them up to 100 ms$^{-1}$, and with vertical acceleration that can be five times that due to gravity, an observation in a turbulent flow that could be related to intermittency, as for example in the modeling of the formation of rain droplets (Shaw and Oncley 2001).

We therefore proceed to the analysis of rotating flows with an emphasis on helicity and intermittency. Section \ref{s:heli} explores the nonlinear dynamics of helical flows, Section \ref{s:intermi} discusses the structures that emerge in  strongly rotating turbulent flows and their intermittency properties, and finally Section \ref{s:conclu} is the conclusion.

\section{Nonlinear dynamics of helical flows} \label{s:heli}
\subsection{The conservation of helicity} \label{ss:cons}

One of the most important and useful principles of physics is that of conservation laws linked, through the theorem of Emma Noether (1918), to invariance properties of the underlying equations; indeed, energy conservation, corresponding to invariance through translation in time, led Pauli (and later, Fermi) to hypothesize the existence of neutrinos. Similarly, the conservation of linear and of angular momentum is associated respectively with invariance of the dynamical equations under spatial translation and rotation. Angular momentum conservation is a key ingredient in the understanding of the relative motion of celestial objects, from accretion disks to planets, stars and galaxies.

Conservation of energy is invoked when explaining the observation of the distribution of excitation among a wide range of scales in a turbulent flow: the nonlinear coupling due to advection, a convolution terms when transformed to Fourier space, leads to the feeding of modes at all the scales available to the system (Kolmogorov 1941; see also Frisch 1995). In the two-dimensional case, the energy flows to large scales as hypothesized by Kraichnan (see Kraichnan and Montgomery 1980 for a review), where friction as is the case for the atmosphere of the Earth will stop the cascade from accumulating on the gravest mode. In three dimensions the cascade of energy is towards small scales and is arrested by dissipative processes. The fact that helicity ($H_V=\left<{\bf u} \cdot \mbox{\boldmath $\omega$}\right>$, the correlation between the velocity and the vorticity) is also conserved was discovered much later than for the energy (Moreau 1961, Moffatt 1969).
Note that in quantum mechanics, helicity is related to the relative direction of the particle's motion and its spin, and lack of mirror symmetry goes under the name of chirality. Helicity is not definite positive, unlike energy; it is a topological invariant, representing the degree of knottedness of vortex lines as well as the twisting of vortex lines, and it is a pseudo-scalar: its sign depends on the frame of reference, either right-handed or left-handed (the symmetry group related to its conservation is discussed in Yahalom 1995).  Helicity in turbulent flows can lead to drag reduction, and to better mixing of chemical components in helical coherent structures (Duquenne et al. 1993,  Zimmerman 1996). Helicity has been measured in the atmosphere (Anthes 1982,  Davies-Jones 1984, Rotunno 1984, Lilly 1986, Markovski et al. 1998, Lewellen and Lewellen 2007) and has been invoked to explain the long life-time of tornadoes and super-cell storms because of weakened non-linearities when it is strong.

The lack of definite sign of helicity has interesting consequences: one sign of helicity, say at small-scale, can be dissipated and yet represent a source of helicity at large-scale of the opposite sign. This renders the interpretation of helicity dynamics more complex. The invariance of $H_V$ in the absence of kinematic viscosity $\nu$ appears rather clearly when writing the Navier-Stokes equations in terms of the Lamb vector ${\cal L}={\bf u} \times \mbox{\boldmath $\omega$}$ with ${\cal P}=p+|{\bf u}|^2/2$ a modified pressure:
\begin{equation}
\frac{\partial {\bf u}}{\partial t} + \mbox{\boldmath $\omega$} \times
  {\bf u} + 2 \mbox{\boldmath $\Omega$} \times {\bf u}  = 
  - \nabla {\cal P} + \nu \nabla^2 {\bf u} + {\bf F}  \ ;
\label{eq:momentum} \end{equation}
mass conservation reduces to $\nabla \cdot {\bf u} =0$ assuming incompressibility for simplicity with a unit constant density, and ${\bf F}$ is an external mechanical force that drives the turbulence, mimicking for example a convective input of energy. In the presence of an imposed solid body rotation with frequency $\Omega$, the rotation axis will be chosen to be in the $z$ direction: $\mbox{\boldmath $\Omega$} = \Omega \hat{z}$; finally, note that ${\cal P}$ is modified by the centrifugal term in the presence of rotation. The global amount of helicity in the flow can be modulated through the forcing term (or in the absence of forcing, by the initial conditions); forcing (and/or initial conditions) can be taken with random phases with a prescribed Fourier spectrum, and with a prescribed amount of helicity through the coupling of two random vectors (see e.g., Pouquet and Patterson 1978). Alternatively, some large-scale order, emanating from an instability, can be prescribed around a given scale $2\pi/k_0$, e.g., through well-known flows like the Taylor-Green flow (TG hereafter) that mimics the laboratory experiments in fluids between two counter-rotating cylinders (see e.g. Monchaux et al. 2007) or Beltrami flows such as the ABC (Arnold 1972). In the case of the TG flow, global helicity is zero because of symmetries although there are strong fluctuations of the local helicity density, whereas Beltrami flows have their velocity and vorticity parallel (or anti-parallel) globally. Moreover, Beltrami flows are known to be unstable (Arnold 1972; Kraichnan, 1973), and their dynamics is chaotic (H\'enon 1966), making them good candidates for the dynamo problem of generation of magnetic fields by turbulent flows (Arnold and Korkina 1983, Galloway and Frisch 1984; Gilbert 1991, Galloway and Proctor 1992) including at small magnetic Prandtl number (Mininni 2007). 

How helical  a flow is can be evaluated with the relative helicity corresponding to the degree of alignment of the velocity and vorticity, i.e. $H_V/(|{\bf u}| |\mbox{\boldmath $\omega$}|)$; it is defined in Fourier space as,  with $ r(k)\le 1$ because of Schwarz inequality:
\begin{equation}
  r(k)=|H(k)|/kE(k) \ .
\label{relative} \end{equation}
Turbulent fluids are known to develop helical structures which are persistent since their associated non-linear advection (the Lamb vector) is weak with ${\bf u}$ and $\mbox{\boldmath $\omega$}$ almost aligned (Pelz et al. 1985, Moffatt 1983, Moffatt and Tsinober 1992,  Holm and Kerr 2002); thus, the temporal evolution of helical flow structures at the onset of the dissipative range where vorticity is strongest takes place on the slow dissipative time scale. Moreover, even when the total helicity is negligible, local helicity density is produced in the flow (Sanada 1993). This overall tendency toward alignment in the flow is fast (Matthaeus et al. 2008), occurring in a turn-over time $\tau_{NL}=L_0/U_0$ with $U_0$ and $L_0$ a characteristic velocity and length scale for the flow; it can be related to the alignment of pressure gradients and shear and has been observed both in DNS and in solar wind data. Even though small-scale structures are found to be strongly helical (of either sign), the helicity of a flow does not seem to alter its dynamics: indeed, evidence stemming from two-point closures of turbulence and from direct numerical simulations of isotropic, and homogeneous turbulence performed on incompressible fluids and with periodic boundary conditions, indicate (both for weak or strong global helicity in terms of $r(k)$) that the distribution of energy among scales follows a power-law (Kolmogorov 1941) which, expressed in terms of correlation functions or of structure functions of second order, reads $S_2(\ell)\sim \ell^{2/3}$, with $S_p(\ell)= \left<[u_L({\bf r}+\mbox{\boldmath $\ell$})-u_L({\bf r})]^p\right>$ the $p$th-order longitudinal structure function on a distance $\ell$, $u_L$ being the projection of the velocity field along the vector $\mbox{\boldmath $\ell$}$. Similarly, it was shown using the renormalization group that the helical contribution to eddy viscosity is sub-dominant, with a $r^{-4}$ dependence in the limit $r \rightarrow \infty$, whereas the renormalized viscosity follows a  classical $r^{-2}$ diffusion law (Pouquet et al. 1978).

\subsection{The ideal case} \label{ss:ideal}

Kraichnan (1973) derived the statistical ensemble equilibria that emerge in three dimensions in the absence of viscosity for a system with a finite number of modes in terms of its global energy and helicity. Defining Fourier spectra such that $H_V=\left<{\bf u} \cdot \mbox{\boldmath $\omega$} \right>=\int H(k){\rm d}k$ and $E_V=\left<|{\bf u}|^2/2\right> = \int E(k){\rm d}k$, the equilibria are:
\begin{equation}
  E(k)=\frac{4 \pi k^2}{\alpha (1-\chi^2)} 
  \ , \ \hspace{0.2cm}
  H(k)= \frac{8 \pi k^4\beta}{\alpha^2 (1-\chi^2)} 
    \ ,\ \hspace{0.2cm}
      \chi = \frac{ k\beta} {\alpha} 
  \  \ , \label{HelSpec} \end{equation}
with $\alpha>0$ and $\beta k_{max}<\alpha$, $k_{max}$ being the maximum wavenumber of the truncated system; this condition is to ensure integrability, namely the positivity of energy, and $|H(k)| = 2|\chi| = 2k|\beta|/\alpha \le kE(k)$. 
 The values of the Lagrange multipliers $\alpha$ and $\beta$ are linked to the two invariants, $E_V$ and $H_V$. In the non-helical case ($\beta=0$), one recovers equipartition of energy among modes which, in 3D, leads to a $k^2$ energy spectrum. Equations (\ref{HelSpec}) show that there is no accumulation of energy or helicity at small wavenumber, unlike the case of 2D-NS. The relative helicity grows with wavenumber although for $k_{max}\rightarrow \infty$ and $\alpha > \beta k_{max}$ finite (corresponding to finite energy), one must have $\beta \rightarrow 0$.

It has long been argued that inviscid dynamics is an indicator of what happens in the dissipative case. It is on the basis of ideal spectra peaking at large scale in 2D-NS that the inverse cascade of energy was postulated by Kraichnan. For three-dimensional Euler taking into account helicity, no such inverse cascade takes place in the non-rotating case. It was also shown recently that similarly to the non-helical case, inviscid dynamics leads to transient energy and helicity cascades that agree with the helical absolute equilibrium given above in Eq. (\ref{HelSpec}): the excess of relative helicity found at small scales in the viscous run is interpreted as a thermalization (Krstulovic et al. 2009), and the large-scale spectra can be recovered with a model using a scale-dependent eddy viscosity. The ideal case in the presence of rotation is being studied presently.

\subsection{Dimensional scaling of Fourier spectra with or without rotation} \label{ss:various}
 
 \begin{table}
\caption{Some of the possible types of cascades for fluid turbulence based on phenomenological arguments, taking into account the presence or absence of either helicity or rotation; $\tau_{tr}$ is the transfer time and $\Omega$ the imposed rotation. See text for details.}
\begin{ruledtabular}  \begin{tabular}{cccc}
\hline

$\downarrow$Type / $\tau_{tr}\rightarrow$   &  $\tau_{E}$, $\Omega \equiv 0$  & $\tau_{H}$, $\Omega \equiv 0$   & $\tau_E^2\Omega$  \\
\hline
\hline
        &  {\bf K41}      & tHcE      & {\bf D92} \\ 
Energy  & e=h=5/3         & 2e+h=4  & e=h=2       \\ 
cascade& a=2/3, b=0 &3a=7-3e,  3b=3e-5 & a=1/2=f=g, b=0 \\ 
  & c=-1/3, d=1  &3c=2(3e-4),  d=-2b & c=-1/2, d=1   \\ 
\hline
$k_D$ & $[\epsilon_E/\nu^3]^{1/4}$ &  & $[\epsilon_E/\nu^2 \Omega]^{1/2}$ \\
Dual ?   & yes     & for e=h=4/3: K04    & yes       \\
\hline \hline
  &  B73 & K04     & {\bf M09}      \\
Helicity   & e+2h=5          & e=h=4/3 & e+h=4, a+c=0, \\ 
cascade &3a=7-3e, 3b=3e-5
  & a=1, b=-1/3 & 3a+3b=2-f , b+d=1, \\ 
  &6c=3e-7, 6d=11-3e & c=0, d=2/3     &3b=3e-5-2f, f+g=1   \\
\hline
$k_D$   &  $[\epsilon_E^a \epsilon_H^b/\nu^2]^{1/[1+e]}$   & 
  $[\epsilon_H/\nu^3]^{1/5}$    
& $[\epsilon_E^a \epsilon_H^b/\nu \Omega^g]^{1/[e-1]}$ \\
  Dual ?   & for e=h=5/3: K41 & yes    & for e=h=2: D92   \\
  \hline
\end{tabular}  \end{ruledtabular}
\label{tab1}
\end{table}

In the presence of viscosity, several spectral dynamics have been envisaged for helical flows. When generalizing them to include the effect of rotation, these possibilities are summarized in Table \ref{tab1}, including some cases not derived previously in the literature. The various regimes are obtained assuming either (or both):
\begin{itemize}
\item Constancy of the flux of energy $\epsilon_E\equiv dE_V/dt$. 
\item Constancy of the flux of helicity $\epsilon_H\equiv dH_V/dt$.
\end{itemize}
For simplicity, anisotropic effects are omitted (they can be added in a straightforward manner). The following timescales are also used, with $\tau_E$ and $\tau_H$ the turn-over times associated with energy and helicity (see below):
\begin{itemize}
\item A characteristic time of the inertial waves in rotating cases as $\tau_w\sim 1/\Omega$.
\item  A cascade time of energy and helicity to small scales $\tau_{tr}$ that can depend on a combination of $\tau_E$,  $\tau_H$ and $\tau_w$.
\end{itemize}

Note that the following type of arguments can be extended to other types of waves, such as for stratified flows, or in magnetohydrodynamics,  for Alfv\'en waves and whistler waves (see Galtier et al. 2005 for specific examples).

Whereas the cascade time $\tau_{tr}$ is generally thought to be the eddy turn-over time $\tau_E\sim [k^3E(k)]^{-1/2}$ in the classical Kolmogorov cascade, one could envisage a characteristic time for transfer based on helical dynamics, $\tau_H\sim  [k^2H(k)]^{-1/2}$ (Kurien et al. 2004), with $\tau_E=\tau_H$ in the case of maximal helicity, or alternatively a time that incorporates in some fashion the dynamical effects of waves. In that latter case, a phenomenological argument due to Iroshnikov (1963) and Kraichnan (1965) stipulated, in the context of Alfv\'en waves in MHD, that the transfer time can be evaluated as $\tau_{tr}\sim \tau_E/{\bar {\epsilon}}$ with ${\bar {\epsilon}}=\tau_w/\tau_E$ the small parameter of the problem: in weak turbulence, the waves are assumed to be fast compared to the nonlinear coupling of eddies. This is in agreement with the fact that nonlinear dynamics is slowed down and weaker in the presence of waves; furthermore, the spectra evaluated in this manner coincide with those found in weak turbulence theory (Zakharov et al. 1992, Newell et al. 2001) when neglecting the effect of anisotropy (see Galtier et al. 2000 for MHD, and Galtier 2003 for rotation).

\begin{figure} \begin{center}     \includegraphics[width=12cm]{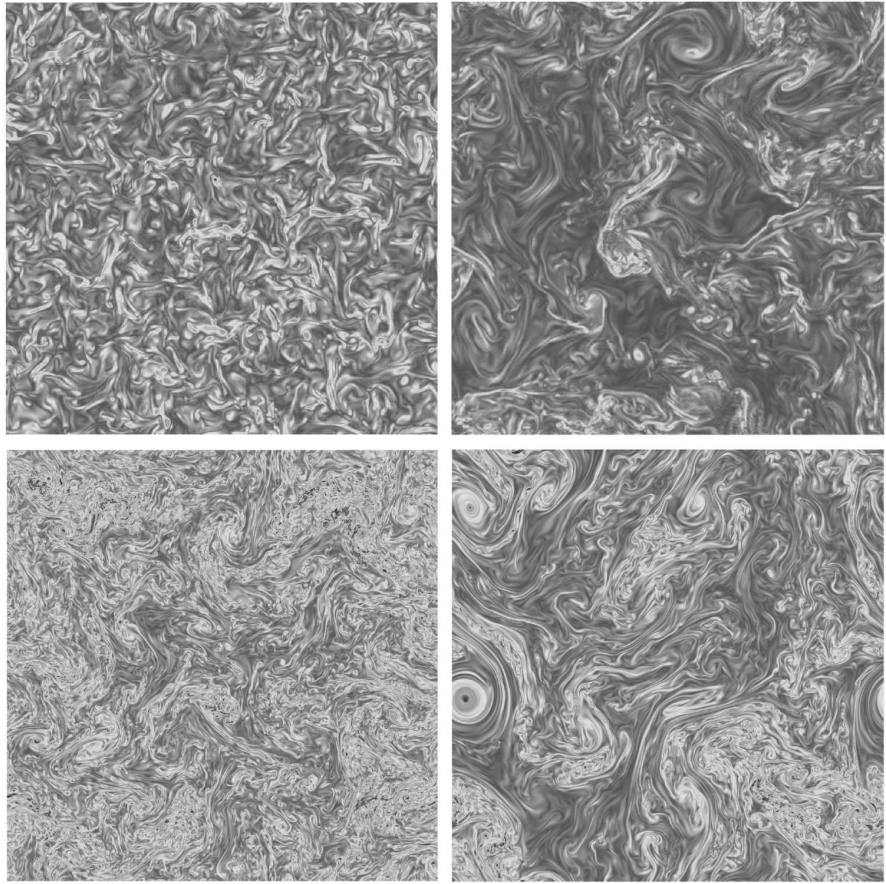}
\caption{
Slices of vorticity in the $xy$ plane in simulations of turbulence in periodic boxes. 
{\it Top left:} $512^3$ simulation of non-helical non-rotating turbulence, Reynolds number $Re=1100$. 
{\it Top right:} same with $\Omega=8$ (Rossby number $Ro=0.07$). 
{\it Bottom left:} $1536^3$ simulation of helical turbulence at early times with $\Omega=9$ ($Re=5100$ and $Ro=0.06$). 
{\it Bottom right:} same at late times. Note the development of strong and smooth vorticity in the latter case, identified as columnar `` Beltrami Core Vortices''. In the top right plot, vortices associated to columns are barely observable, being less organized and with more small scale features.
} \end{center} \label{fig:slice} \end{figure}

\begin{figure} \begin{center} \includegraphics[width=14cm]{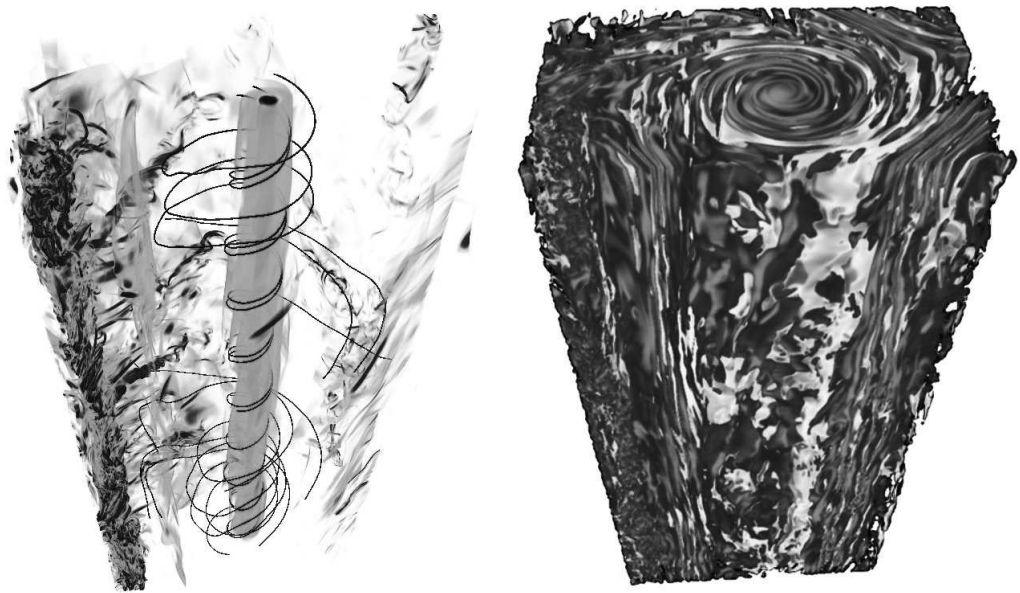}
\caption{
{\it Left:} Three dimensional rendering of vorticity using VAPOR (Clyne et al. 2007) showing the spatial juxtaposition of a Beltrami Core Vortex (BCV) and a vortex tangle, for the same run at late times as shown in Fig. 1 (bottom right). The width of the core is approximately $1/7$ the size of the computational domain. Superimposed on the BCV are fluid particles trajectories which, in the vicinity of these structures, are laminar and helicoidal. 
{\it Right:} visualization at the same time of velocity-vorticity alignment. One can observe the core of the structure with positive (dark) relative helicity and concentric rings of alternate sign of helicity (dark and light) surrounding it, whereas, in the vortex tangle, no clear structure emerges.
} \end{center} \label{fig:PVR_W_HR} \end{figure}

\begin{figure} \begin{center} \includegraphics[width=12cm]{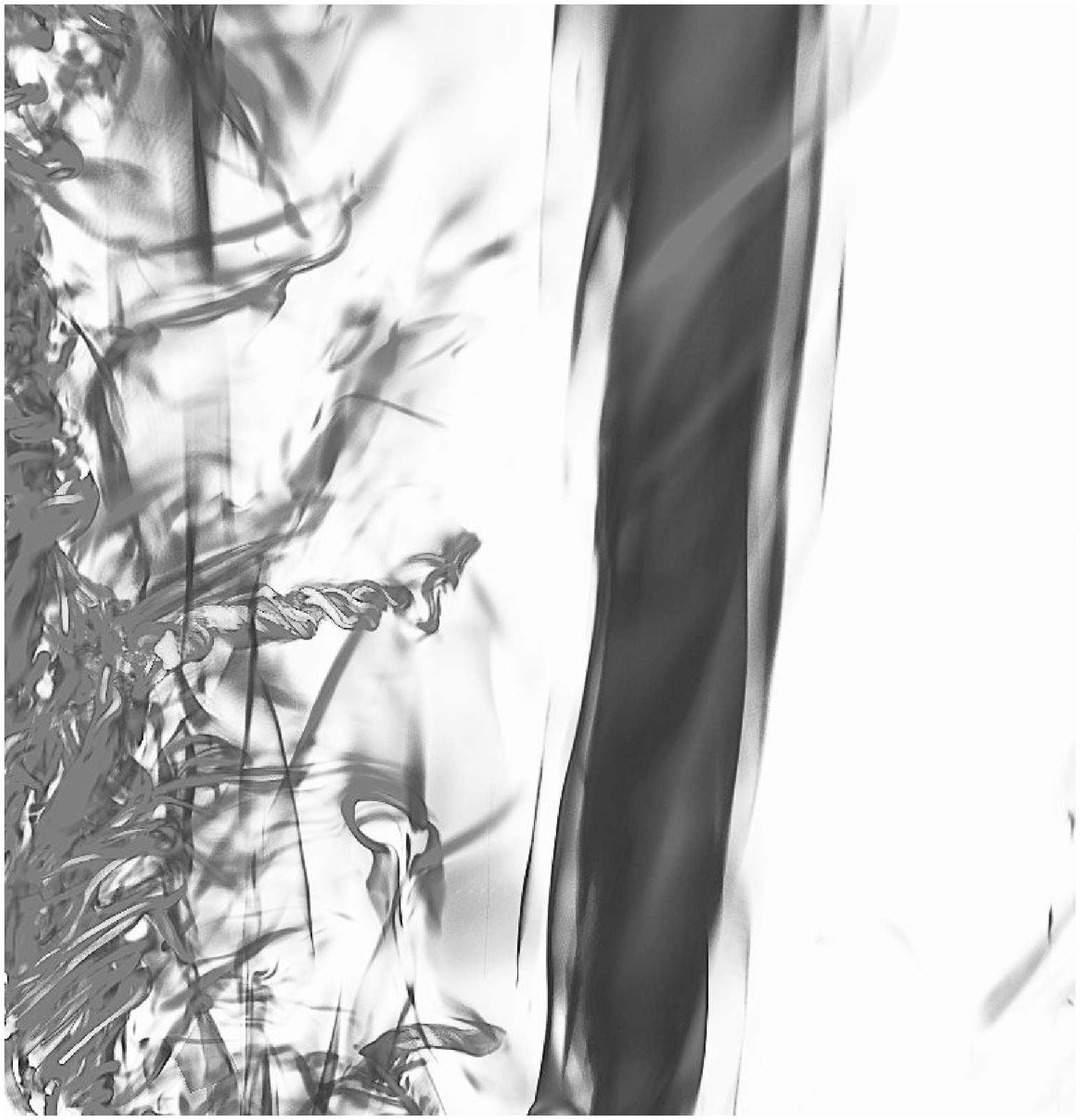} 
\caption{
Zoom on vorticity intensity, when strong, at late times in the same simulation as in Fig. 2. 
The co-location of laminar structures, ``Beltrami core vortex'' (BCV) and of a tangle of vortex filaments with more complex paths and at smaller scale, to the left of the BCV, is clear. The BCV columns, different from Taylor columns (see text) are updrafts cyclonic vortices that are fully helical (see Fig. 2) and thus live for a long time, whereas the small-scale vortices are more evanescent, living for a couple of eddy turn-over times. Note two less-intense and 
inter-twined smaller multiple vertical vortex cores to the left. 
} \end{center} \label{fig:PVR_W_zoom} \end{figure}

Specifically, what is assumed is that the energy and helicity spectra take the form:
\begin{equation}
 E(k)=C_E \epsilon_E^a  \epsilon_H^b \Omega^f k^{-e}  \ , \ \ \ \ \ \ \ \ \ \ 
 H(k)=C_H \epsilon_E^c  \epsilon_H^d \Omega^g k^{-h}  \ ,
\label{eq:spectra}\end{equation}
with $\epsilon_E$ and $  \epsilon_H$ the energy and helicity flux rates respectively, $C_E$ a generalized Kolmogorov constant and $C_H$ the corresponding constant for the helicity spectrum, and with $\Omega$ the rotation frequency in the presence of an imposed solid body rotation as stated before. This choice covers the physics we examine in this paper and most cases studied in the literature, with energy, helicity and rotation all taken into account; the eight indices $[a,b,c,d,e,f,g,h]$ are the exponents to be determined through a combination of dimensional analysis and phenomenology, with $e,h$ the spectral indices.

The results of this analysis are summarized in Table \ref{tab1}; in its  nomenclature, K41 stands for the classical Kolmogorov (1941) phenomenology, extended to a joint energy-helicity cascade; D92  stands for the energy cascade mediated by inertial waves (Dubrulle and Valdettaro 1992; Zhou 1995) and for its new extension to the helical case. B73 is for the dual cascade of energy and helicity as spelled out in Brissaud et al. (1973), including the case of zero energy flux. K04 stands for the case of a cascade fashioned by the helicity time-scale (Kurien et al. 2004). The case of a helicity cascade predominant over energy and mediated by rotation is M09 (Mininni and Pouquet 2009{\it ab}). Finally, the generalization to the case where the time-scale based on helicity is the relevant feature of the cascade is dealt with in case tHcE  without rotation (the similar case with rotation is omitted for simplicity but yields $e+h=3$). In bold-face ({\bf K41}, {\bf D92} and {\bf M09}) are indicated cases that have been observed in the laboratory, in atmospheric flows or in direct numerical simulations; note that for a $k^{-4/3}$ helicity cascade (K04), the DNS observation concerns the so-called bottleneck effect between the classical Kolmogorov range and the dissipation range, an effect which may also be attributed to incomplete thermalization (Frisch et al. 2008) or prominence of non-local interactions near the cut-off wavenumber; $k_D$ is the dissipation wavenumber computed assuming equality of the viscous time $1/[\nu k^2]$ and the relevant transfer time $\tau_{tr}$ to small scales (either $\tau_E$, $\tau_H$ or 
$\tau_{NL}^2/\tau_w=\tau_E^2 \Omega$) at that wavenumber. 

The direction of cascades is not mentioned in Table \ref{tab1}; it is well-known that the energy can be cascaded towards either the small scales (standard three-dimensional case in the absence of rotation) or to the large scales (2D-NS case). It has also been observed to undergo both a direct and an inverse cascade in the presence of rotation (Smith et al. 1996, Mininni et al. 2009, Mininni and Pouquet 2009a).
Finally, by ``dual'' cascade is meant an energy cascade with a constant flux simultaneously with a helicity cascade with its own constant flux; this implies, using $\epsilon_E\sim kE(k)/\tau_{tr}$ and $\epsilon_H\sim kH(k)/\tau_{tr}$, the relationship 
$$
H(k) = E(k) \frac{\epsilon_H}{\epsilon_E}
$$
which in turn implies $e=h$, $f=g$, $c=a-1$ and $d=b+1$ in Equ. (\ref{eq:spectra}) for this dual cascade.

In more detail, the type of reasoning behind the estimations for the energy and helicity spectra listed above and detailed in Table \ref{tab1} is very simple: on the one hand, one argues that, by definition, an inertial range has constant flux, independent of wavenumber. A flux is the ratio of the total energy or helicity, divided by a characteristic time which we associate to the transfer time $\tau_{tr}$. There are several candidates for $\tau_{tr}$ as discussed before. We eliminate the dissipation time which is supposed to be long compared to other relevant times in the problem, by definition of a high Reynolds number flow. 

So the transfer time can be the eddy turn-over time $\tau_{E}\sim \ell/u_{\ell}$; this classical choice leads, in three dimensions, to the Kolmogorov energy spectrum and the helicity in this scenario follows the energy cascade with a constant flux as well.
In the absence of rotation, that is actually the only solution that is observed: a dual Kolmogorov law (to within intermittency corrections), {\it viz.}
$$
 E(k)=C_E\epsilon_E^{2/3}k^{-5/3} \ , \ \ \ \  H(k)=C_H \epsilon_E^{-1/3} 
 \epsilon_H k^{-5/3} \ .
$$
This result of a dual Kolmogorov cascade was obtained using two point closures of turbulence (Andr\'e and Lesieur 1977), DNS (see e.g. Chen et al. 2003b and references therein) and the early phases of ideal dynamics  (Krstulovic et al. 2009). Note that, whereas the variation of $C_E$ with Reynolds number has been documented with a slow convergence with $Re$ (Ishihara et al. 2005), the corresponding   variation of $C_H$ with Reynolds number, or with different flows, is unknown at this stage and difficult to ascertain in the laboratory because of the difficulty to measure helicity. Furthermore, whereas for the ideal case, $r(k)\sim k$ (see Eq. (\ref{HelSpec})), one finds numerically that in the dissipative case $r(k)\sim 1/k$, i.e. indicating a recovery of mirror-symmetry in the small scales, albeit at a slow ($1/k$) rate. Furthermore, although the maximal condition $r(k)=1$ is not observed to be attained globally in turbulent flows, the alignment between velocity and vorticity is known to emerge in small-scale structures, namely in vortex filaments (Moffatt and Tsinober 1992, Matthaeus et al. 2008).

Dropping the assumption of a constant helicity flux, one can find other solutions on dimensional grounds, namely $c=4/3-h$ and $d=h-2/3$. If instead of $\tau_E$ a transfer time based on helicity is used to regulate the cascade, other solutions can also be obtained. In the presence of helicity, a time that differs from $\tau_{E}$ can be constructed using the sweeping time associated with the helicity spectrum, namely $\tau_H\sim [\ell/(u_{\ell}\omega_{\ell})]^{1/2}$; this leads to a $k^{-4/3}$ range, both for the helicity and for the energy (Kurien et al. 2004); in this latter case, one finds $E(k)\sim \epsilon_E \epsilon_H^{-1/3} k^{-4/3}$ and $H(k)\sim  \epsilon_H^{2/3} k^{-4/3}$ (case K04 in Table \ref{tab1}).

As mentioned before, in the presence of waves a new characteristic time has to be taken into account as when $\tau_w\le \tau_E$ nonlinear interactions leading to a cascade of energy to small scales are damped. Weak turbulence theory (the small parameter of the problem that allows for closures of the equations being the ratio $\bar \epsilon=\tau_w/\tau_E$) leads to a set of integro-differential equations in terms of the various spectra of the problem. These complex equations can be shown to have both zero-flux (statistical equilibria) and constant flux (turbulent) solutions in terms of power laws of the wavenumber (for ``warm' cascades that combine zero-flux and constant flux solutions for the energy, see Connaughton and Nazarenko 2004). These weak turbulence solutions can in fact be recovered using a simple phenomenological argument which incorporates in a straightforward manner the small parameter $\bar \epsilon$ in the problem. Namely, one says that the transfer to small scales is slowed down as modeled by a longer transfer time $\tau_{tr}\sim \tau_E/ \bar \epsilon$. Taking $\tau_w\sim 1/\Omega$ for inertial waves, one obtains $\tau_{tr}\sim \tau_E^2/\tau_w \sim \Omega/[k^3E(k)]$. Writing now that the transfer of energy to small scale is evaluated locally as $\epsilon_E \sim k^4E^2(k)/\Omega$ leads immediately to $E(k)\sim [\epsilon_E \Omega]^{1/2} k^{-2}$. In order to obtain the dependency of the spectrum as written in Eq. (\ref{eq:spectra}), one uses dimensional analysis which yields $a=f=1/2$ (Dubrulle and Valdettaro 1992; Zhou 1995). 
The generalization to the case of a dual energy-helicity cascade is straightforward and gives for the helicity spectrum $h=2, g=1/2, c=-1/2, d=1$ as is given in Table \ref{tab1} (case D92); this results in:
$$
 E(k)=C_E[\epsilon_E \Omega]^{1/2}k^{-2} \ , \ \ \ \  H(k)=C_H  [ \Omega/\epsilon_E]^{1/2} \epsilon_H   k^{-2}    \ ;
$$
 the anisotropic version reads $ E(k_{\perp},k_{\parallel})\sim [\epsilon_E \Omega k_{\parallel}^0]^{1/2}k_{\perp}^{-5/2}$ and 
 $H(k_{\perp},k_{\parallel})\sim [\epsilon_H/\epsilon]E(k_{\perp},k_{\parallel})$, assuming that most of the transfer is in the perpendicular direction and that $k_{\parallel}^0$ stems from initial conditions (one recovers isotropy with $k_{\parallel}^0\sim k_{\parallel}\sim k_{\perp}$). Note that this formulation of the spectra ensures that the dissipation wavenumber for the energy and helicity spectra is the same, namely $[\epsilon_E/\nu^2 \Omega]^{1/2}$ (see Table \ref{tab1}).
 
What if one were to use a higher power of the small parameter $\bar \epsilon=\tau_w/\tau_E$ in the evaluation of the transfer time? Let us take for example $\tau_{tr,2}\sim \tau_E /\bar \epsilon^2$. An analysis similar to what has been done above easily leads to $E(k)\sim \epsilon_E^{2/5}\Omega^{4/5}k^{-11/5}$, as proposed by Zeman (1994). The difference in terms of power law of the energy spectrum is small and given that intermittency may steepen the spectra, it will be hard to distinguish the Zeman spectrum from the D92 spectrum, {\it viz.} $k^{-11/5}$ {\it versus} $k^{-2}$. However, note that, by assumption, the dependency on the imposed rotation rate is significantly higher for the Zeman case, a point that may be amenable to examination using DNS and a parametric study; again, the generalization to a dual cascade gives $h=-4/5$ for the spectral index of helicity (note that this case is not incorporated in Table \ref{tab1}).

\subsection{The helicity-dominated cascade to small scales} \label{ss:helical}

\begin{figure} \begin{center}     
\includegraphics[width=6.17cm]{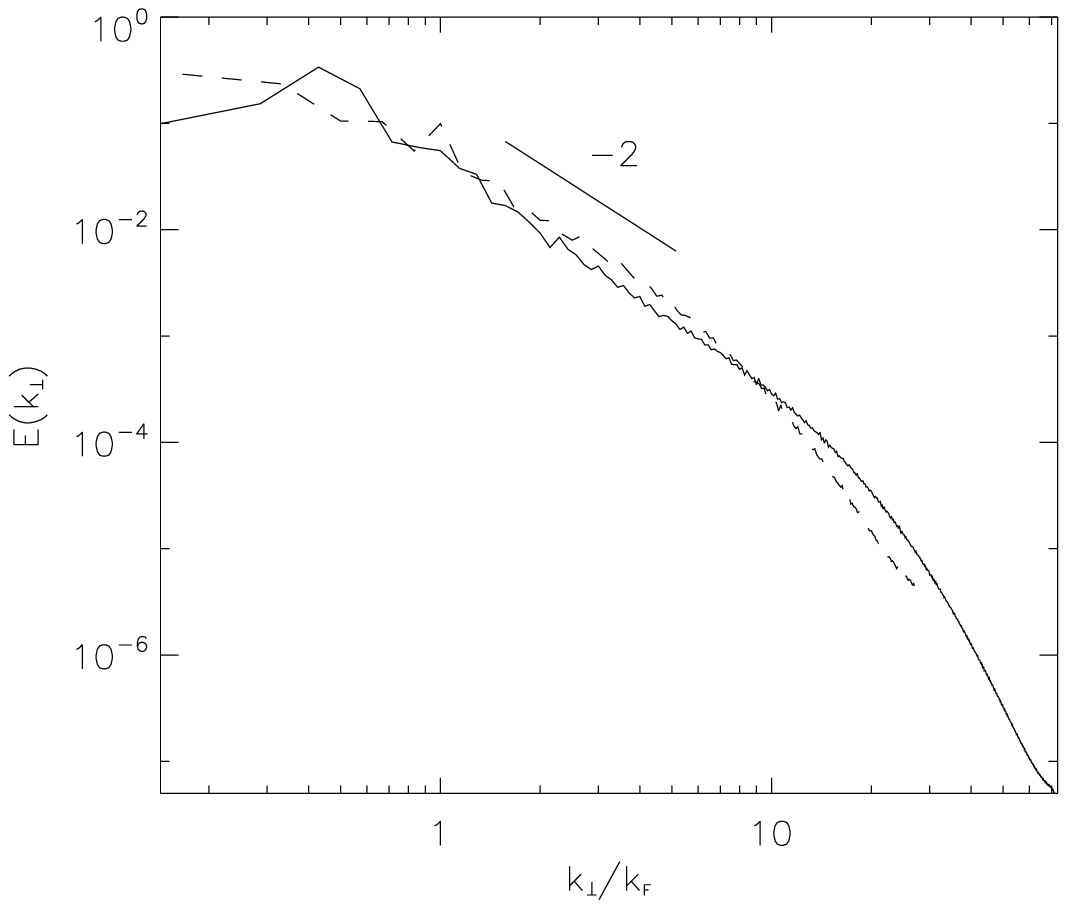}
\includegraphics[width=6.17cm]{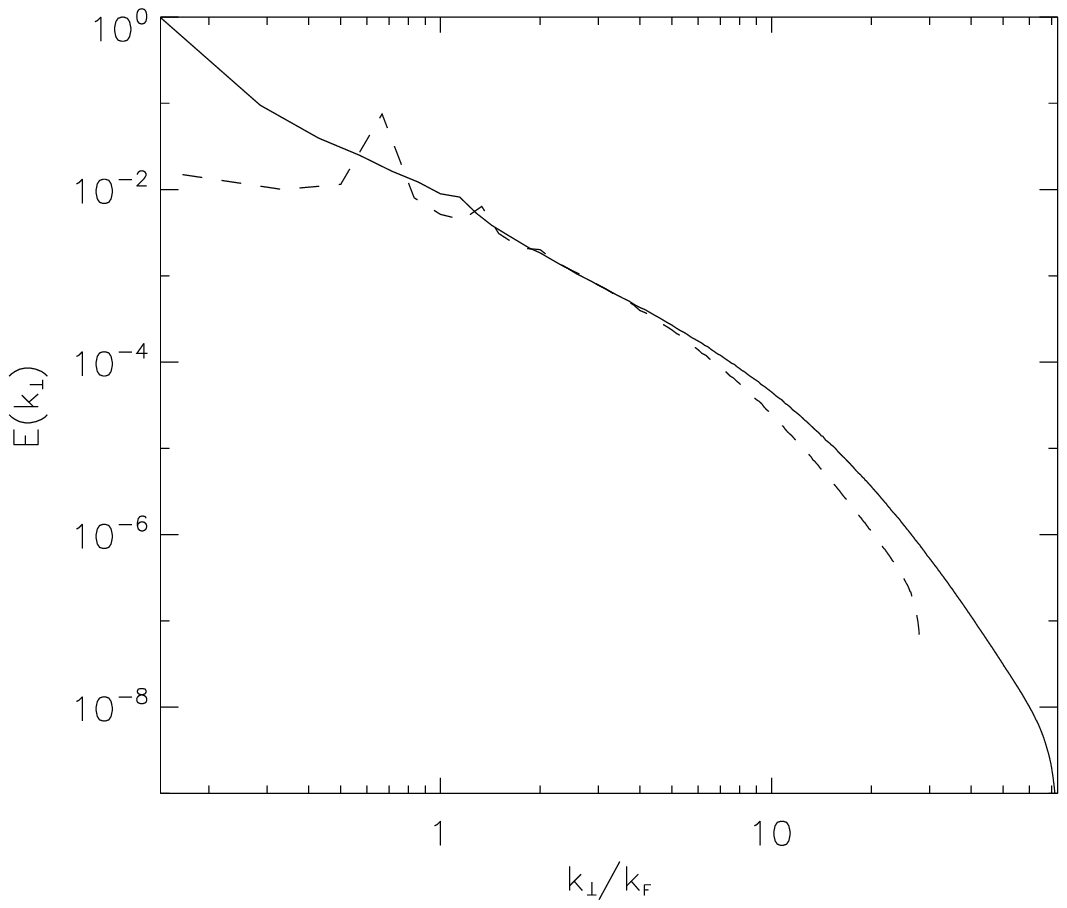}
\caption{
{\it Left:} perpendicular energy spectrum in a simulation of helical rotating turbulence with $Re=5100$ and $Ro=0.06$ (solid), and in a simulation of non-helical rotating turbulence with $Re=1100$ and $Ro=0.07$ (dashed). The straight line indicates a power law $\sim k^{-2}$. 
{\it Right:} parallel energy spectrum for the same runs. In both figures the wavenumbers are normalized by the forcing wavenumber $k_F$.}
\end{center} \label{fig:spec} \end{figure}

One can make another type of hypothesis, supposing that, for some reason the energy transfer becomes negligible compared to the helicity transfer to small scales. This could be justified, e.g.,  on the basis that in the presence of strong rotation, energy flows to large scales and thus only a small amount of energy is available for a direct cascade to small scales. One is then faced solely with a helicity cascade, as done in Brissaud et al. 1973 (see case B73 in Table \ref{tab1}). This leaves to an indetermination in the spectral indices of $E(k)$ and $H(k)$, with $e+2h=5$, simply because the eddy turn-over time is expressed in terms of $E(k)$ and the constant flux of helicity is in terms of $H(k)$. This solution is compatible with the classical Kolmogorov case ($e=h=5/3$) when both fluxes become comparable; it also admits $e=7/3$, $h=4/3$ which is the only solution with no dependence on $\epsilon_E$ but only on $\epsilon_H$; it is also maximal ($h=e-1$).  A pure helical cascade $\sim k^{-7/3}$ is compatible with the exact law written in Gomez et al. (2000) based on the conservation of helicity, when assuming maximal helicity leading to dimensional scaling ($\omega_{\ell}\sim u_{\ell}/{\ell}$ at scale $\ell$) and assuming further no correlation between the velocity and vorticity fields. The difference of scaling in the presence of helicity may be related to the effect of large-scale flows (Olla 1998), as is observed for wall turbulence and the Bolgiano-Obukhov scaling for stratified flows; however, the lack of evidence for a $-7/3$ law in laboratory experiments and DNS is strengthened by the use of nearest-neighbor shell models of turbulence (Olla 1998).

Although we are not aware of any observation relative to the $e+2h=5$ law put forward by Brissaud et al. (1973), the similar solution in the case of rotating flows has been observed in numerical simulations. The one important difference with the case of Brissaud et al. (1973) is that now the transfer of energy to small scales is mediated by waves and with a characteristic time taken to be $\tau_{tr}\sim \tau_E^2/\tau_w$. This leads now to a new constraint on the energy and helicity spectral indices that read 
$$e+h=4 \ ,$$ 
compatible with the dual cascade previously known ($e=h=2$, see D92 in Table \ref{tab1}) but possibly leading to different spectra and thus to a loss of universality in helical rotating turbulence. The solution with $e\not= h$ is clearly observed in several DNS (Mininni and Pouquet (2009 {\it ab}) and in Large-Eddy Simulations as well (Baerenzung et al. 2008, 2009). Why do we observe this interesting solution in the presence of rotation and not the corresponding $e+2h=5$ in its absence? The reason could be quite simple: at small Rossby number, the flow tends to a quasi two-dimensional state and thus supports an inverse energy cascade to large scales; the energy flux to small scale is thus negligible compared to the helicity flux to small scales, in accordance with the hypothesis of the models in this section. Note however that the energy that is transferred to small scales cannot be zero since $E(k)\ge H(k)/k$; evaluating this inequality at the smallest resolved scale of the flow, for example at the dissipation length $k_D$, indicates that the condition on the energy spectrum becomes smaller the higher the wavenumber available to the system, i.e. the higher the Reynolds number at a given rotation rate. Similarly, at a fixed Reynolds number, the higher rotation rate ensures more wave-dominated regime that leads to a more clear inverse energy cascade. This is consistent with results from a parametric study using a spectral model of small-scales that incorporates the effect of helicity on transport coefficients (Baerenzung et al. 2009).

One can finally note that one may find a bit odd that, in a dual cascade, the sub-dominant field has in its expression a dependency on both $\epsilon_{E}$ and $\epsilon_{H}$. This is consistent, though, with the fact that finite dissipation must result for both spectra and that the dissipation is evaluated at $k_D$ whose expressions depend on the transfer time assumed to be relevant to the particular problem.
 
\begin{figure} \begin{center}     
\includegraphics[width=6.17cm]{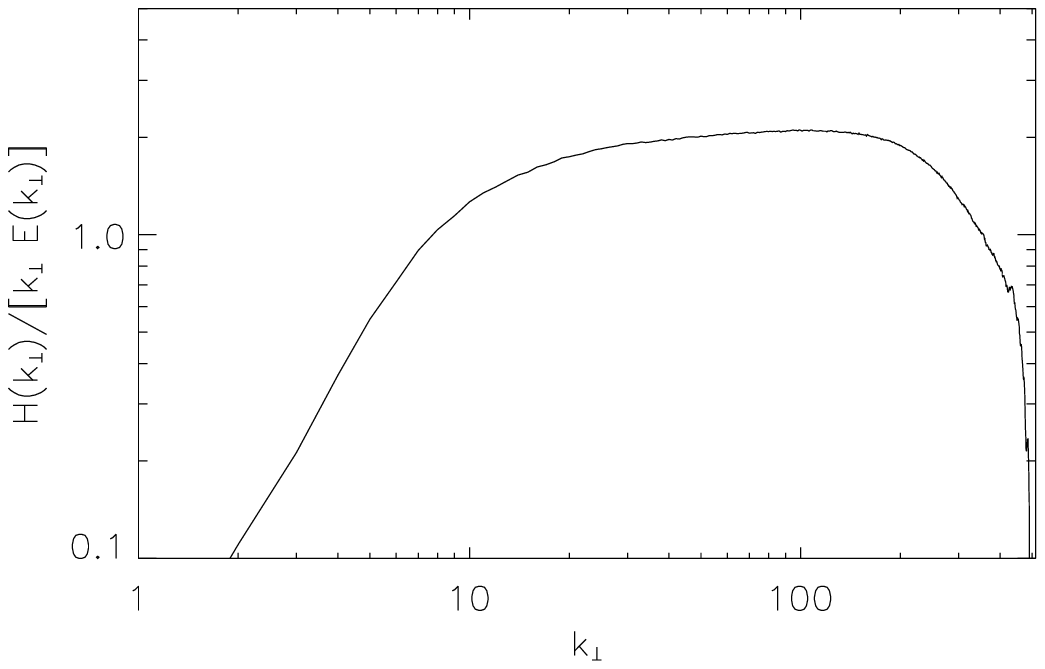}
\includegraphics[width=6.17cm]{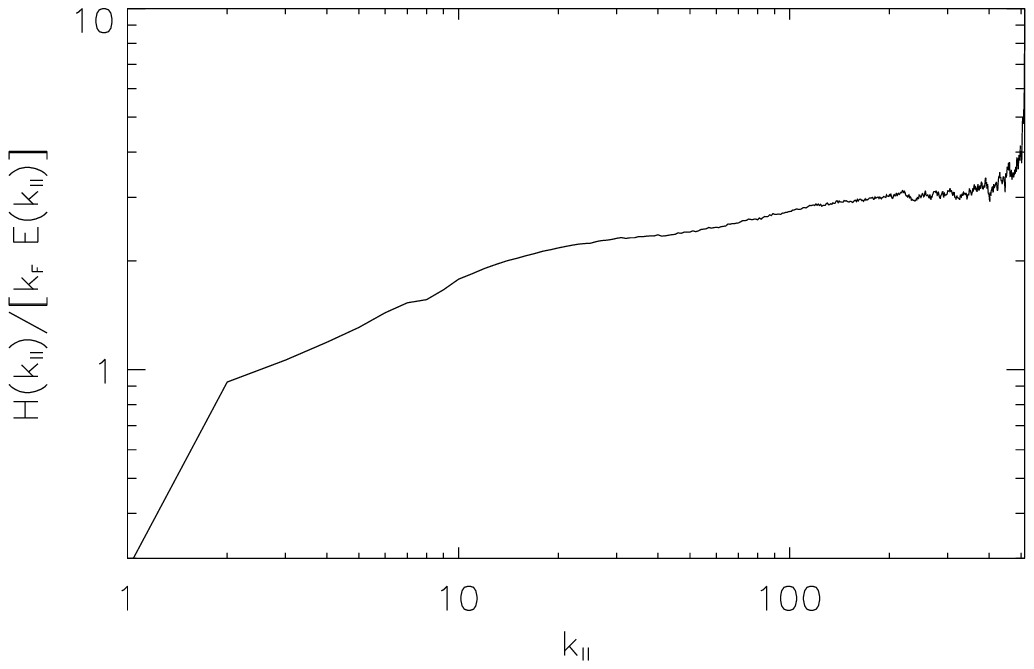}
\caption{Relative helicity in terms of the perpendicular (left) and parallel (right) wavenumbers in a $1536^3$ simulation of helical rotating turbulence with $Re = 5100$ and $Ro =0.06$.}
\end{center} \label{fig:relh} \end{figure}

\subsection{A remark on polarized helical waves} \label{ss:polarization}

Several further comments concerning the analysis presented in Table \ref{tab1} are in order. First, the constancy of the two (energy and helicity) cascades put sufficient constraints on the system for allowing a determination of most exponents, with some mild assumptions; otherwise, if only one cascade is hypothesized (say, the energy), a power law can occur in the spectrum of the other field (here, the helicity) that does not correspond to a constant flux solution for the helicity. Second, in some cases it may be difficult to define a power law solution for a non-definite positive field like the helicity, which may undergo changes of sign in its inertial range. A similar problem arises in MHD when looking at the cascade of the ideal invariant $H_C=\left<{\bf u} \cdot {\bf b}\right>$. Using second-order two-point closures of turbulence, it was shown that in fact the cross-helicity $H_C$, after a transient, establishes a two-lobe spectrum of one sign at large scale and of the opposite sign at scales smaller than the dissipation length (Grappin et al. 1982). A similar analysis has not been performed for helicity in three-dimensional fluids and this remains an open question that can be tackled using both closure equations (Andr\'e and Lesieur 1977) and DNS. However, when forcing the flow with Beltrami waves, the sign of helicity is predominantly that of the imposed forcing and thus one has no more difficulty evaluating the inertial index of the helicity spectrum in that case than for the energy. Furthermore, in MHD $H_C$ can be seen as the difference of two (positive) energies, using the identity $4{\bf u} \cdot {\bf b} = |{\bf u} + {\bf b}|^2 - |{\bf u} - {\bf b}|^2$, and its cascade can reflect subtle compensations between these two (pseudo-)energies $|{\bf u}\pm {\bf b}|^2$. Similarly, helicity can be expressed in terms of the difference of two energies. Indeed, the complexity of the behavior of helical flows can be better understood when considering simultaneously the $[E,H]$ and the $[E^{\pm}, H^{\pm}]$ behavior of spectra where $E^{\pm}$ and $H^{\pm}=kE^{\pm}$ refer to the eigenfunctions of the curl operator, corresponding to left-hand and right-hand circularly polarized maximally helical waves (Kraichnan 1973; Herring 1974; Waleffe 1993). Following Ditlevsen and Giuliani (2001) (see also Olla 1998), one can identify in the helical case a scale $\ell_H$ at which dissipation sets in by simply writing a balance between input and dissipation: 
$$\epsilon_H \sim \nu \int^{K_H} k^2 H(k) dk  \sim \nu u \omega/\ell^2_H \ ,$$
where the upper bound in the integral is the dissipation wavenumber associated with the helicity spectrum, $K_H=2\pi/\ell_H$; using a dimensional estimate $\omega \sim u/\ell$ and  the Kolmogorov scaling law $u\sim \epsilon_E^{1/3}\ell^{1/3}$, one arrives at 
$$\ell_H\sim [\nu^3 \epsilon_E^2/  \epsilon_H ^3]^{1/7} \ ,$$
a scale which is larger than the dissipative Kolmogorov length scale $\ell_D\sim [\epsilon_E/\nu^3]^{1/4}$; this would imply that the helicity spectrum begins its dissipative range before the energy range in Fourier space. Note however that, if instead of supposing maximal helicity as  done here, one simply expresses the dissipation in terms of the helicity spectrum
$$H(k)\sim \epsilon_E^{-1/3}   \epsilon_H k^{-5/3}$$
then one arrives at $K_H=K_D$. The statement $K_H/K_D \le 1$ is a point contradicted by numerical experiments (see e.g., Chen et al. 2003b, Alexakis et al. 2006). A remarkable explanation was given in Chen et al. (2003b) where one can find a detailed analysis of the interactions between all these modes based on the energy and helicity spectra as well as the spectra of the helical polarized waves: the scale $\ell_H$ determines the arresting of the spectra of the $\pm$ variables but the energy and helicity starts to dissipate at $\ell_D=2\pi/k_D$, the difference being attributed to cancellations between the fluxes of the $\pm$ waves. 
These considerations are related to the fact that, for maximal helicity, one would have $H(k)\sim k E(k) \sim k^{-2/3}$, leading to divergences of its dissipation. It is a peculiar property of the helical wave decomposition that they are indeed fully helical ($H^{\pm}=kE^{\pm}$) with $H=H^+-H^-$: helicity and its dissipation thus remain finite due to cancellations between the fluxes (Chen et al. 2003b). The fact that the dissipation scale for the helicity is equal to $\ell_D$ is consistent with the idea of a balance between energy input and decay with $H(k)\sim \epsilon_E^{-1/3}  \epsilon_H k^{-5/3}$, as noted above. The scale $\ell_H$ enters in the helical wave decomposition to render the maximal helicity $H^{\pm}$ constant. The dynamics of the $[E,H]$ fields and the $[E^{\pm}, H^{\pm}]$ fields thus differ, and the cancellations between the polarized waves occur in such a way that the helicity cascade is slaved to that of the energy. This remark also confirms the analysis in Kraichnan (1973) in terms of coupling of helical waves: a maximal helicity state is not consistent with the nonlinear dynamics even if in the initial state there is maximal helicity; in other words, a maximal helicity state assuming a Kolmogorov spectrum $E(k)\sim k^{-5/3}$ leads to a non-physical helicity spectrum, which is only realized and observed in the $\pm$ variables.

\section{Structures  and intermittency} \label{s:intermi}
\subsection{The emergence of Beltrami Core Vortices} \label{ss:BCV}

The preceding considerations rely on dimensional analysis on the classical basis of constant-flux solutions to the turbulence problem and using different time scales. Of all the solutions envisaged in Table \ref{tab1}, only some have been observed, as stated before, and more work exploring fully parameter space remains to be done, with different Reynolds and Rossby numbers and different forcing functions. However, spectra are only a simple and constrained way to examine the data. What about the characteristic structures that emerge in rotating flows, and their statistical properties?

We thus now report on some of the features of rotating turbulence obtained through a massive direct numerical simulation on a grid of $1536^3  \approx 3.6\times 10^9$ points and forced with the ABC flow, with $L_F=2\pi/k_F$ the characteristic scale of the forcing, of amplitude $F_0$:
\begin{eqnarray}
{\bf F} &=& F_0 \left\{ \left[B \cos(k_F y) +   C \sin(k_F z) \right] \hat{x} + \right. {} \nonumber \\
&& {} + \left[C \cos(k_F z) + A \sin(k_F x) \right] \hat{y} + 
   {} \nonumber \\
&& {} + \left. \left[A \cos(k_F x) + B \sin(k_F y) \right]   \hat{z} \right\}.
\label{eq:ABC} \end{eqnarray}

Results concerning the overall dynamics of the flow and its intermittency properties are reported in detail in Mininni and Pouquet (2009{\it ab}). The Navier-Stokes equations in a rotating frame  (see Eq. (\ref{eq:momentum})) are integrated with a pseudo-spectral code and periodic boundary conditions using a second-order Runge-Kutta temporal scheme. The flow is first led to establish a statistically steady state with $\Omega=0.06$, i.e. in the near absence of rotation, a phase taking  roughly ten turn-over times; with $F_0=0.5$ and $k_F=7$, the resulting rms velocity is $U\approx 1$ and the Reynolds number
$$R_V=UL_F/\nu \approx 5600$$
with the choice of $\nu=1.6\times 10^{-4}$. Then, at a time labeled $t=0$ in the following, the rotation is set to $\Omega=9$, corresponding to a Rossby number
$$Ro= U/(2\Omega L_F)\approx 0.06 .$$
The time step is $\Delta t=2.5\times10^{-4}$. Note that the value of the forcing scale is chosen so as to let both a direct cascade to small scale and an inverse cascade to large scale develop simultaneously. In so doing, this run can be viewed as a combination of two computations performed earlier (Mininni and Pouquet 2009a) at lower resolution (with $512^3$ grid points instead of $1536^3$ here), in which either the direct cascade (with $k_F=2$) or the inverse cascade (with $k_F=7$) were studied separately. The computation is then performed for 30 turn-over times $\tau_{NL}=L_F/U$, corresponding to 180 inertial wave periods. At early times, the dynamics is dominated by the Coriolis force and by resonant interactions between inertial waves. Because the dispersion relation is anisotropic and favors $k_{\parallel}=0$, the flow does not develop any substantial variation in the vertical and, in accordance with the dynamic Taylor-Proudman theorem, it displays columnar structures first clearly observed in the laboratory by Hopfinger et al. (1982) and numerically by several authors (see e.g. Bartello et al. 1994, Cambon and Scott 1999, Smith and Waleffe 1999). Note that the maximal vorticity in this flow is on average roughly 50 times its {\it rms} value and peaks are found up to 70 times that, whereas in a similar flow without rotation the maximal vorticity averages 20 times its {\it rms} value with peaks up to 30 times that value. The larger ratio of fluctuations to {\it rms} value in the vorticity of the run with rotation is associated to a substantial decrease in the enstrophy and the energy dissipation rates when rotation is strong. The maximum value of vorticity in the volume is however only weakly dependent on rotation, indicating that turbulence and small scale structures still develop in the flow even at small Rossby number provided the Reynolds number is large enough, and provided one waits long enough in terms of the rotation period.

Figure 1 shows four horizontal slices of the vorticity intensity for a turbulent flow without helicity nor rotation (top left), a rotating flow without helicity at late times ($t\approx 40$, top right); a rotating flow with helicity at early times ($t\approx 6.5$, bottom left), and a rotating flow with helicity at late times ($t\approx 30$, bottom right). The vorticity of a rotating flow without helicity at early times is not show as it looks similar to the non-rotating case. As a result, the vorticity in both helical and non-helical rotating flows at early times looks similar. As time evolves and anisotropy develops, the rotating flows tend to develop column-like structures in the velocity and vorticity (see e.g., Smith and Waleffe 1999). It is important to note, however, that such structures are different in the helical and in the non-helical case. While the columns in the non-helical flow have small scale structure and display a myriad of vortex filaments, in the helical case a few columns develop a core with a smooth helical velocity field. One of these structures can be observed on the left ( bottom-right slice) in Fig. 1.

When visualizing the vorticity magnitude in three dimensions in a subset of the total volume for the helical rotating flow (see Fig. 2), using Perspective Volume Rendering as implemented with the VAPOR software (Clyne et al. 2007), these laminar structures can be fully appreciated. Particle trajectories around the column are helicoidal, as materialized by the dark lines; the cores are fully helical and we thus name them Beltrami core vortices (BCV) to distinguish them from Taylor columns. These BCV are updrafts and cyclonic structures that live for long times (they were tracked for over ten turn-over times in the simulation) because their associated Lamb vector is weak; they support Kelvin waves along them, and move around because they are embedded in the far-field velocity. Their number decreases with time through reconnections because of the tendency of the flow to form larger-scale structures as time elapses in an inverse cascade. Overall, the vorticity is strongest in the tangle of vortex filaments, whereas the vertical velocity is coherent and strongest in the laminar columns (not shown). The vortex core of the column is surrounded by a calm region with weak vorticity (note the emptiness of vortical structures in the surroundings of the column) which acts as a transition region between the laminar and the turbulent flow. Far from these structures, column-like structures with a tangle of small scale vortex filaments also develop, as in the non-helical rotating case (see the dark column on the left of the subvolume in Fig. 2). The BCV are Beltrami globally whereas beltramization in the tangle of vortices is local and random (see the relative helicity for the same region in Fig. 2).

The origin of these stable structures can be identified  by integrating backward in time the particle trajectories: they correspond to regions of large helicity where the columns form. Note that a rigorous scaling for the velocity profile in the coherent vortices that emerge at late times in 2D-NS is found to be $u(r)\sim r^{-5/4}$ (Chertkov et al. 2009). How this scaling is altered by rotation is being presently investigated (Lebedev, private communication), and whether such scaling obtains in any of the coherent vortices that are found in DNS of rotating flows, either at early or at late times is an open topic for future research.
The complex array of smaller-scales vortices next to the BCV is shown in more detail in Fig. 3, reminiscent of the vorticity field observed in many turbulent flows without rotation, except that one does perceive organization in the vertical direction in the form of a large-scale column, as is also found in the case of non-helical rotating fluids when the Reynolds number is large enough (Mininni et al. 2009). These columns have a more complex helicity structure and disappear on a time of the order of the eddy turn-over time. In the relative helicity of both structures shown in Fig. 2 (right) one can clearly distinguish the large-scale order associated with the BCV, with helicity of one sign, and a succession of rings of opposite signs (alternate light and dark regions), whereas as soon as one approaches the complex vortex tangle at smaller scale, no large-scale organization can be identified.

The tangle of vortex filaments surrounding a laminar structure, together with more complex, larger and spiraling features, is reminiscent of observations of multiple core vortex tornadoes. Of course, only the very basic ingredients of helical convective storms and tornadoes are present in our computation: the input of energy that mimics the convective instability, the helicity often observed (Lilly 1986), and the strong rotation due to the local environment (Rotunno 1984). If obviously many other features of such an extreme event are absent in this bare-bone model (such as moisture, boundary effects, or micro-physical processes in general), there remains the possibility that the stability of these meteorological phenomena may be linked to the intrinsic dynamics of a rotating helical turbulent flow. The onset and ensuing acceleration of the tornadic motions can only be related to its environment and would require a more complete model; for example, it is already known that the growth of large-scale helical structures is observed in the presence of convection (Levina and Burylov 2006).

The laminar organization of the velocity and vorticity fields is not observed in helical isotropic and homogeneous turbulence, nor is it observed in rotating flows without helicity as shown for example in Fig. 1. At early times, the bi-dimensionalization of the flow leads to columnar structures that are not surrounded by vortex tangles, the latter being due to the nonlinear terms coming into play. The interplay between rotation (which breaks the mirror-symmetry in the evolution equations) and helicity (which quantifies departures from mirror-symmetry of the flow) is the driving agent for the formation of strong localized and persistent columnar structures, even though helicity is mostly transferred towards smaller scales (Mininni and Pouquet 2009{\it ab}) and is itself strongly intermittent.

Finally, the next two figures show traditional statistics of turbulent flows, namely the energy spectrum (Fig. 4) reduced into its parallel and perpendicular components (i.e., in terms of variation of wavenumber parallel and perpendicular to the rotation axis), and the relative helicity spectra (Fig. 5) also separated in parallel and perpendicular dependence. In the perpendicular energy spectrum the inverse cascade can be observed as energy piles up at wavenumbers smaller than the forcing wavenumber. Noteworthy is the fact that a clear scaling at scales smaller than the forcing appears in the energy in terms of $k_{\perp}$, with the non-helical case close to $\sim k_\perp^{-2}$ scaling (see e.g., Jacquin 1990, Zeman 1994, Zhou 1995, and Cambon et al. 2004) and the helical case showing a slightly steeper spectrum (see Mininni and Pouquet 2009{\it ab}). However, no clear scaling is observed in the parallel wavenumbers. In the helical run, an excess of relative helicity at small scales is observed in both the vertical and the horizontal directions (see Fig. 5).

\subsection{Intermittency and self-similarity in rotating flows} \label{ss:p/2}

\begin{figure} \begin{center}     
\includegraphics[width=11cm]{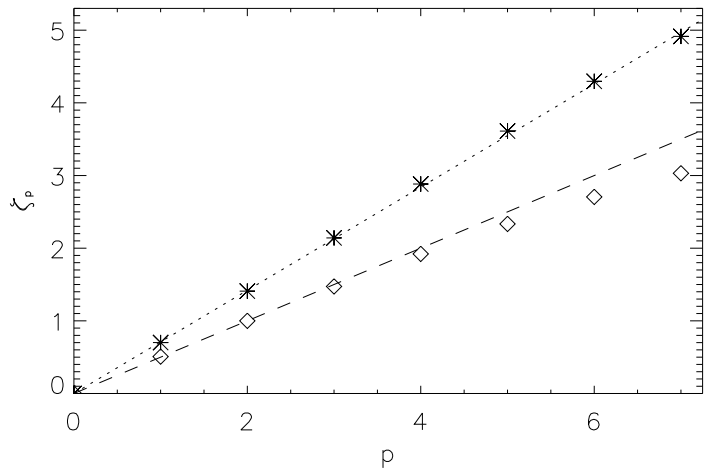}
\caption{Scaling exponents $\zeta_p$ in the direct cascade range of helical (stars) and non-helical (diamonds) rotating turbulence from direct numerical simulations. The dotted line corresponds to $\zeta_p=0.71 p$ and the dashed line to $\zeta_p=p/2$. Note the deviation of the diamonds from the straight line for large values of $p$ (see also Mininni and Pouquet 2009c).}
\end{center} \label{fig:expo} \end{figure}

Turbulent flows, in three dimensions and in the absence of rotation, display non-Gaussian statistics in their small-scales, associated with strong localized structures such as vortex filaments. How does the presence at small scales of laminar columns in rotating helical flows alter such properties?
Specifically, intermittency is measured through scaling exponents of structure functions; for the velocity field, with $u_L$ the velocity component projected onto the distance ${\bf r}$, one can examine:
$$<\delta u_L^p(r)>\sim r^{\zeta_p} \ ,$$
assuming isotropy and homogeneity, and with the increment of a function $f$ defined as $\delta f(r)=f(x+r)-f(x)$. In a rotating flow, one can also introduce increments on distances ${\bf r}_\perp$ perpendicular to the axis of rotation and measure
$$<\delta u_L^p(r_\perp)>\sim r_\perp^{\zeta_p^{\prime}} \ .$$
When $\zeta_p=ap$, one talks of complete self-similarity of the flow with, in the case of the isotropic and homogeneous Kolmogorov (1941) energy spectrum, $a=1/3$. Departures from such a linear scaling indicates that the flow is multi-fractal with a suite of exponents defining its small-scale properties and presumably characteristic of localized and strong intermittent structures; such departures can be computed exactly in the framework of a model of the passive scalar, but for three-dimensional Navier-Stokes turbulence, there is no such theory although some models are quite close to the data, e.g. the She-L\'ev\^eque model (1994).
The observed curvature of the $\zeta_p=f(p)$ law is related to the fact that the probability distribution functions of velocity gradients have fat tails, with strong departure from Gaussianity at high values, corresponding to strong structures that are concentrated in space within the flow.  No such departure from a linear law is observed for laminar flows, and similarly none obtains in inverse cascades toward large scales, as mentioned previously.

In the presence of helicity, one may ask what are the intermittent properties of the flow in the absence of rotation. This was studied in Chen et al. (2003a) where they defined intermittency of helicity on its flux to small scales; they found that the intermittency of the velocity is comparable to that in the non-helical case, but that the intermittency of the helicity is larger than for the energy, insofar as the scaling exponents of high-order structure functions depart further from the linear scaling $\zeta_p=p/3$, a result somewhat reminiscent of the case of the passive scalar.

In the presence of rotation, several laboratory experiments and DNS have addressed this issue as well. New experiments using a non-intrusive (electromagnetic) forcing have allowed for detailed statistics of rotating flows using SPIV (Stereoscopic PIV) up to $R_{\lambda}\sim 240$ and micro Rossby numbers of 0.15 (van Bokhoven et al. 2009). At low intensity of the forcing, self-similarity of the flow seems to obtain, with $\zeta_p\sim p/2$, getting close to $\zeta_p\sim3 p/4$ as the intensity of the forcing current is increased. Other experiments reported near $\zeta_2 \sim 1$ scaling (see e.g., Simand et al. 2000, and Baroud et al. 2002 for experiments with stirring), although small deviations from the self-similar scaling $\zeta_p \sim p/2$ were also reported. As a result, whether rotation makes the flow self-similar or just decreases the intermittency is a matter of debate. In numerical simulations of forced rotating turbulence, $\zeta_p \sim p/2$ with small intermittency departures (when compared with non-rotating turbulence) were found using non-helical forcing (M\"uller and Thiele 2007, Mininni et al. 2008). However, in the case of helical forcing at very large resolution, one obtains  $\zeta_p \sim 0.71 p$, and furthermore no deviations from the self-similar straight line was found within the error bars (Mininni and Pouquet 2009 {\it ac}). In the numerical simulations it was also found that the exponents get closer to the straight line as time evolves. Note that when self-similarity is observed, with $\zeta_p=ap$, then a normalization of intermittency exponents by the data at some order $n$ as in the ESS (extended self similarity hypothesis) methodology (Benzi et al. 1993), gives trivially $\zeta_p/\zeta_n=p/n$. 

For decaying flows, different intermittency exponents were observed as the flow decays (Morize et al. 2005, Seiwert et al. 2008), from classical non-rotating values at small times evolving toward $\zeta_p\sim p/2$ as one approaches times of the order of $1/\Omega$, although strict self-similarity was not observed (the values found for the exponents in these experiments are consistent within error bars with the values reported in the numerical simulations of M\"uller and Thiele 2007, and Mininni et al. 2008). For very long times in the experiments, $\zeta_2 \ge 2$, in contradiction with a $E(k)\sim k^{-2}$ law (D92 of Table \ref{tab1}), but in agreement with the law put forward in Mininni and Pouquet (2009a) that emphasizes the role helicity plays in the dynamics of a rotating flow; indeed, a strictly non-helical flow (implying orthogonality of the velocity and vorticity everywhere in space) is hard to obtain, even if, because of symmetries, the global helicity may be close to zero, as in the Taylor Green flow (Monchaux et al. 2007).

Figure 6 illustrates these results showing the scaling exponents from two simulations of forced rotating turbulence with and without helicity. The exponents for the non-helical flow correspond to $t\approx 40$ ($\Omega=8$, and therefore $t \Omega \approx 320$). While $\zeta_2 \approx 1$, weak deviations from the self-similar scaling are observed for $p \le 5$. For the helical flow the exponents were measured at $t\approx 30$ ($\Omega=8$, and $t \Omega \approx 270$) and all exponents are consistent with $\zeta_p \sim p0.71$ within error bars.

Thus a possible explanation for energy spectra found to be steeper than $k^{-2}$ in some of the experiments (e.g., Seiwert et al. 2008), and for the different behavior reported for the high order exponents, in some cases consistent with intermittency and in others with self-similarity, may be that some helicity is present in the flow, sufficiently so to affect the overall dynamics; this would be an interesting point to check both experimentally and numerically, in the latter case by varying the relative helicity of the forcing and observing the scaling that obtains. Another possible explanation may be related to some dependence of the exponents with the Rossby number, or with time. This point does raise the questions (see e.g., Seiwert et al. 2008, Baerenzung et al. 2009) of whether the transition from a Kolmogorov state to a steeper power law is smooth as the Rossby number is decreased, and whether for fixed Rossby number there is an asymptotic scaling the system reaches for very long times.

In the case of helical rotating flows, the coexistence of laminar columnar Beltrami Core Vortices with a complex vorticity field at smaller scale must have an impact on the statistics of the velocity field; it is natural to try to associate the self-similar energy cascade illustrated in Fig. 6 to the dynamics of the BCVs whereas the helicity cascade (which is intermittent) may be associated with the small-scale vortex tangle; to that effect, a study using wavelets that allow to examine both the scale variation and the space variation of structures is in progress.

The confirmation of self-similarity in turbulence in the combined presence of helicity and rotation (conditions that are relevant to many atmospheric flows) would relate the dynamics of such three-dimensional flows to the advances made in two-dimensional turbulence and critical phenomena in general. However, to use renormalization group techniques (Ma and Mazenko 1975, Forster et al. 1976) a small parameter needs to be identified, besides the Rossby number that governs the energetic exchanges between turbulent eddies and waves when dealing, e.g., with the weak turbulence regime (Zakharov et al. 1992,  Connaughton et al. 2003, Nazarenko and Schekochihin 2009; see also Cambon et al. 2004, for a clarification of the link between two-point closures and weak turbulence). Indeed, the smallness of the ratio of the inertial wave period to the eddy turn-over time has already been used to derive integro-differential equations in terms of energy and helicity spectra (Galtier 2003) in the context of weak turbulence. However, the weak turbulence solutions are not observed in the numerical studies or in many atmospheric flows for at least three reasons: ({\it i}) the numerical resolution may be insufficient to see such laws; ({\it ii}) the theory is non-uniform in scale and the weak turbulence limit breaks down; ({\it iii}) in the case of rotation the inverse cascade of energy is not present at lowest order in the theory and thus the solution selected by this approach is one of an energy cascade to small scales, whereas the numerical data of helical rotating turbulence indicates that this cascade is sub-dominant to the helicity cascade (Mininni and Pouquet 2009b). The candidate (backed by numerical data) for a small parameter in this problem is the (adimensionalized) ratio $\chi=\epsilon_E/ L_F \epsilon_H$, where $\epsilon_E$ and $\epsilon_H$ are the direct energy and helicity fluxes, constant by definition in the inertial range. The energy flux to small scales is all the more negligible as more energy is transferred to large scales in an inverse cascade. Whether similar phenomena take place under the bi-dimensionalization of a flow due to other external constraints (such as stratification or an imposed magnetic field) is unknown at this point.

\section{Conclusion} \label{s:conclu}

The interplay between waves, nonlinearities and energy/helicity transfer and intermittency is a topic of debate presently for a wide variety of waves (Connaughton et al. 2003) with, in some cases, the determination of power-law behavior in the wings of probability density functions (see, e.g. the discussion in the context of surface gravity waves in the ocean in Choi et al. 2005). However, the quasi-bi dimensionalization of the flow under an external agent such as rotation or a uniform magnetic field renders dimensional analysis more delicate since it introduces anisotropy in the scaling laws, and the presence of helicity has not been taken into account in general in these studies.

The results presented here may indicate ways to study helical rotating turbulence from a theoretical point of view. In particular, we reviewed the many phenomenological scaling laws derived in the literature and in the present paper, based on the different time scales present in the system, considering both rotation and helicity. Some of the scaling laws were reported in experiments or numerical simulations, while others where never observed. This may give us some information on what are the relevant time scales for the dynamics, as well as being a hint to what are the dominant interactions between the waves and eddies. In particular, it is worth pointing out that in isotropic and homogeneous turbulence only the dual cascade of energy and helicity with Kolmogorov scaling has been observed so far (except for intermittency corrections and the bottleneck effect at the onset of the dissipation range), a solution which is obtained using the eddy turnover time and constant fluxes of both energy and helicity. In the rotating case more solutions have been reported, although all have the direct energy flux substantially decreased (and the time scale of the cascade increased) as a result of interactions with waves which select the resonances.

Recent numerical simulations are consistent with different scaling laws in the direct cascade range of helical and non-helical rotating turbulence. We briefly compared these two cases and discussed the phenomenological arguments that are consistent with the observed spectra. Finally, we discussed recent experimental and numerical studies of intermittency in rotating flows. The conflicting results about intermittency (in some cases supporting intermittency, although decreased by rotation, and in others indicating self-similarity) may be associated to the effect of helicity in the flow, or to dependence of the intermittency exponents with the Rossby number or with time. However, when comparing numerical simulations, it was found that two runs at similar Rossby number and at similar times (albeit at different Reynolds number) display self-similar behavior or decreased intermittency depending on whether the flow had helicity or not. It is unclear for the moment whether the scaling exponents of the non-helical flow will behave as the helical ones for smaller values of the Rossby number or for later times.

Self-similarity of these flows would open new possibilities of theoretical developments. It may lead the way to theoretical progress in unraveling the structure of turbulent flows, with possible extensions to the study of hairpin vortices in turbulent mixing layers and boundary layers (Rogers and Moin 1987) such as the planetary boundary layer, where stratification will play an important role as well. The finding of multi-scale structures that coexist spanning the range from the smallest dissipative scales in the flow to the largest energy containing scales, preserving scale invariance and with a small parameter associated with them, would relate the study of such complex flows to critical phenomena, where solvable models that preserve the complexity of the underlying processes exist. Finally, the self-similarity found at least for helical rotating flows can be exploited by subgrid models (as, e.g., in Baerenzung et al. 2009) which are often based only on second order statistics of the flow, to fruitfully study higher Reynolds numbers, lower Rossby number or larger separation of scales. Scale invariance provides the needed framework for the development of such models and should prove particularly fruitful in such cases. Several sub-grid models can be devised in this context. Lautenschlager et al. (1988) proposed, on the basis of similar analyses performed in the case of coupling to a magnetic field, to add, as parametrization of the small scales, an expression of the form $\alpha \mbox{\boldmath $\omega$ }+ \gamma \Delta \mbox{\boldmath $\omega$}$, with $\alpha$ and $\gamma$ depending on the amount of helicity in the flow, similar to the alpha effect in MHD whereby a large-scale magnetic field is unstable due to the small-scale helicity of the flow. It was shown in Pouquet et al. (1978) using the renormalization group that in fact the $\alpha$ term above is missing when performing a systematic expansion in terms of elimination of the small scales and that, indeed a term proportional to $k^3$ was present, although it was deemed negligible in the limit $k\rightarrow 0$ when compared to the renormalization of the viscosity. In related studies, Frisch et al. (1988) showed that the AKA (anisotropic kinetic alpha) effect can be important for flows lacking parity invariance (invariance under simultaneous reversal of position and velocity), leading to destabilization of the large scales. This may be compared with 
the work of Yokoi and Yoshizawa (1993) for effects in inhomogeneous flows that can arise in the laboratory through imposing a swirling flow to a pipe flow, or in the atmosphere by combining swirling motions with updrafts linked to either convective motions or boundary layer effects.

Finally, we believe several of the recent results suggest that a better understanding of the role of the inverse cascade of energy in rotating flows is required, and is a likely next step in the investigation. How does it coexist with the direct cascade of energy in the non-helical case, and the direct cascades of energy and helicity in the helical case? How does it compare with the two-dimensional case in the absence of rotation, for example in the statistical properties of vortices (see e.g., Macwilliams 1984)? The similarities and differences with the inverse cascade of energy in 2D-NS may help us understand the self-similar behavior, and relate the problem with the recent advances in conformal invariance and two dimensional turbulence (see Bernard et al. 2006 and Cardy et al. 2008).

\vskip0.2truein
{\it 
Computer time provided by NCAR which is sponsored by NSF. PDM is a member of the Carrera del Investigador Cient\'{\i}fico of CONICET.}


\begin{thebibliography} 
{111}

\bibitem{alex}
Alexakis, A., Mininni, P.D. \& Pouquet, A. 2006
Large scale flow effects, energy transfer, and self-similarity in turbulence.
 {\it  Phys. Rev. E} {\bf 74}, 016303.
 
\bibitem{andre_lesieur}
Andr\'e, J.C. \& Lesieur, M. 1977
Influence of Helicity on the  Evolution of Isotropic Turbulence at High Reynolds Number.
{\it J. Fluid Mech.}, {\bf 81}, 187--207.

\bibitem{anthes82}
Anthes, R. 1982 Tropical cyclones, their evolution, structure and effects. {\it Meteorological Monographs} {\bf 19}, Number 41. American Meteorological Society.

\bibitem{arnold}
Arnold, V. I. 1972 Remarks on behavior of the flows of a three-dimensional ideal fluid under a small perturbation of initial velocity field. {\it Pril. Matem i Mekh.} {\bf 36} 255--262 (in russian).

\bibitem{arnold2}
Arnold, V. I. \& Korkina, E.I. 1983 The growth of a magnetic field in a three-dimensional incompressible flow. {\it Vetsnik Moscow State University Ser. Math.} {\bf 3} 43--46 (in russian).

\bibitem{julien_NS}
Baerenzung, J., Politano, H., Ponty, Y., \& Pouquet, A. 2008
Spectral modeling of turbulent flows and the role of helicity. 
{\it Phys. Rev. E} {\bf 77}, 046303.

\bibitem{julien_JAS}
Baerenzung, J., Rosenberg, D., Mininni, P.D. \& Pouquet, A. 2009
Where we observe that helical turbulence prevails over inertial waves
 in forced rotating flows at high Reynolds and low Rossby numbers, in preparation.
 
\bibitem{baroud}
Baroud, C., Plapp, B.,  She, Z-S. \&  Swinney H. 2002
Anomalous Self-Similarity in a Turbulent Rapidly Rotating Fluid.
{\it Phys. Rev. Lett.} {\bf 88}, 114501.

\bibitem{bartello94}
Bartello, P., M\'etais, O. \& Lesieur, M. 1994 Coherent structures in rotating three-dimensional 
turbulence.  {\it J. Fluid Mech.} {\bf 273}, 1--29.

\bibitem{benzi}
Benzi, R. Ciliberto, S., Tripiccione, R.,  Baudet, C., Massaioli, F.  \& Succi, S. 1993 
1993 {\it 	Phys. Rev. E } {\bf 48}, R29-R32. 

\bibitem{bernard06}
Bernard, D., Boffetta, G., Celani, A., \& Falkovich. G. 2006
Conformal invariance in two-dimensional turbulence. {\it Nature Phys.} {\bf 2}, 124--128.

\bibitem{bernard07}
Bernard, D., Boffetta G., Celani, A. \& Falkovich, G. 2007
Inverse turbulent cascades and conformally invariant curves. {\it Phys. Rev. Lett.} {\bf 98}, 024501.

\bibitem{vergassola} 
Boffetta, G., Celani, A. \&  Vergassola, M. 2000
 Inverse energy cascade in two-dimensional turbulence: Deviations from Gaussian behavior. {\it Phys . Rev . E } {\bf 61}, 
 R29--R32.
 
\bibitem{boffeta07}
{Boffetta, G.} 2007
\newblock Energy and enstrophy fluxes in the double cascade of two-dimensional
  turbulence.
\newblock {\em J.\ Fluid Mech.} {\bf 589}, 253--260.

\bibitem{bokh09}
van Bokhoven, L. J. A.,  Clercx, H. J. H.,  van Heijst, G. J. F. and Trieling, R. R., 2009
Experiments on rapidly rotating turbulent flows. {\it Phys. Fluids} {\bf 21}, 096601.

\bibitem{brissaud}
Brissaud, A., Frisch, U., L\'eorat, J. , Lesieur, M. \& Mazure, A. 1973
Helicity cascades in fully developed isotropic turbulence.
{\it Phys. Fluids} {|bf 16}, 1366--1367.

\bibitem{cambon_NJP}
Cambon C., Rubinstein R., and Godeferd, F.S. 2004
Advances in wave turbulence: rapidly rotating flows.
{\it New J. Phys.} {\bf 6}, 73.

\bibitem{scott}
Cambon, C. \& Scott, J.F. 1999 Linear and nonlinear models of anisotropic turbulence.
{\it Ann. Rev. Fluid Mech.} {\bf 31}, 1--53.

\bibitem{cardy_book}
Cardy, J., Falkovich, G.G., \& Gawedski, K. 2008
Non-equilibrium statistical mechanics and turbulence. 
London Mathematical Society, {\it Lecture Note Series} {\bf 355}, 
S. Nazarenko and O. Zaboronski (eds.), London.

\bibitem{chenchen03a}
Chen, Q., Chen, S. and Eyink, G., 2003a
The joint cascade of energy and helicity in three-dimensional turbulence.
{\it Phys. Fluids} {\bf 15}, 361--374.

\bibitem{chenchen03b}
Chen, Q., Chen, S., Eyink, G. and Holm, D., 2003b
Intermittency in the joint cascade of energy and helicity.
{\it Phys. Rev. Lett.} {\bf 90}, 214503.

\bibitem{lebedev}
Chertkov, M. Kolokolov, I. \& Lebedev, V. 2009
Universal Velocity ProÞle for Coherent Vortices in Two-Dimensional Turbulence.
Preprint, see ArXiv.0909.1575.

\bibitem{choi_naza}
Choi, Y., Lvov, Y.V., Nazarenko, S., \& Pokorni, B. 2005
Anomalous probability of large amplitudes in wave turbulence.
{\it Phys. Lett. A} {\bf 339}, 361--369.
 
\bibitem{Clyne07}
{Clyne, J., Mininni, P., Norton, A., \& Rast, M.} 2007
\newblock Interactive desktop analysis of high resolution simulations: application to turbulent plume dynamics and current sheet formation.
\newblock {\em New J.\ Phys.} {\bf 9}, 301.

\bibitem{connaughton_NN_PHYSD}
Connaughton, C., Nazarenko, S., \& Newell, A.C. 2003
Dimensional analysis and weak turbulence.
{\it Physica D} {\bf 184}, 86--97.

\bibitem{connaughton_2}
Connaughton, C. \& Nazarenko, S. 2004
Warm Cascades and Anomalous Scaling in a Diffusion Model of Turbulence. {\it Phys. Rev. Lett.} {\bf 92}, 044501.

\bibitem{davies_jones}
Davies-Jones, R. 1984
Streamwise vorticity: the origin of updraft rotation in supercell storms.
{\it J. Atmos. Sci.} {\bf 41}, 2991--3006.

\bibitem{ditlevsen}
Ditlevsen, P. \&  Giuliani, P. 2001 Dissipation in helical turbulence.
{\it Phys. Fluids} {\bf 13}, 3508--3509.

\bibitem{dubrulle}
Dubrulle, B. \& Valdettaro, L., 1992
Consequences of rotation in energetics of accretion disks.
{\it Astron. Astrophys.} {\bf 263}, 387--400.

\bibitem{chemical}
Duquenne A.M., Guiraud P. \& Bertrand J. 1993
Swirl-induced improvement of turbulent mixing: Laser study in a jet-stirred tubular reactor. 
{\it Chem. Eng. Sc.} {\bf 48}, 3805--3812.

\bibitem{FNS77}
Forster, D., Nelson, D.R. \& Stephen, M. J. 1976
Long-time tails and the large-eddy behavior of a randomly stirred fluid.
{\it Phys. Rev. Lett.} {\bf 36}, 867--870.

\bibitem{frisch_book}
Frisch, U. 1995 {\it Turbulence: The legacy of A.N. Kolmogorov}. Cambridge Univ. Press, Cambridge.

\bibitem{frisch_bottle}
Frisch, U., Kurien, S., Pandit, R., Pauls, W.,  Ray, S., Wirth, A. \& Zhu, J. Z. 2008
Hyperviscosity, Galerkin Truncation, and Bottlenecks in Turbulence.
{\it Phys. Rev. Lett.} {\bf 101}, 144501.

\bibitem{frisch_AKA}
Frisch, U., Scholl, H.,  She, Z-S. \&  Sulem, P.L. 1988
A new large-scale instability in three-dimensional incompressible flows lacking parity invariance. 
{\it Fluid Dyn. Res.} {\bf 3}, 295--298.

\bibitem{galloway1}
Galloway, D. \& Frisch, U. 1984 A numerical investigation of magnetic field generation in a flow with chaotic streamlines.
{\it Geophys. Astroph. Fluid Dyn.} {\bf 29}, 13--18.

\bibitem{galloway}
Galloway, D.J. \& Proctor, M.R.E. 1992
Numerical calculations of fast dynamos in smooth velocity fields with 
realistic diffusion. {\it Nature} {\bf 365}, 691--693.

\bibitem{galtier00}
Galtier, S., Nazarenko, S., Newell, A. \& Pouquet, A. 2000 A weak turbulence theory for incompressible MHD.
{\it J. Plasma Phys.} {\bf 63} 447Ð488. 

\bibitem{Galtier03}
{Galtier, S.} 2003
\newblock Weak inertial-wave turbulence theory.
\newblock {\em Phys.\ Rev.\ E} {\bf 68}, 015301.

\bibitem{galtier05}
Galtier, S.,  Pouquet, A. and Mangeney, A. 2005
On spectal scaling laws for incompressible anisotropic MHD turbulence.
{\it Phys. Plasmas} {\bf 12}, 092310.

\bibitem{ABC} 
Gilbert, A.D. 1991
Fast dynamo action in a steady chaotic flow. 
{\it Nature} {\bf 350}, 483--485.

\bibitem{gomez}
Gomez, T., Politano, H. \& Pouquet, A. 2000
Exact relationship for third-order structure functions in helical flows.
{\it Phys. Rev. Lett.} {\bf 61}, 5321--5325.

\bibitem{grappin}
Grappin, R.,  Frisch, U., L\'eorat, J. \& Pouquet, A. 1982 Alfv\'enic fluctuations
as asymptotic states of MHD turbulence.  {\it Astron. Astrophys.} {\bf 105}, 6--14.

\bibitem{green}
 Greenspan, H.P. 1968 {\it The theory of rotating fluids}, Cambridge University Press. 
 
\bibitem{henon}
H\'enon, M. 1966 Sur la topologie des lignes de courant dans un cas particulier. {\it Comptes Rendus de l'Acad]'emie des Sciences, Paris} {\bf 262}, 312-314 (in french).

\bibitem{herring}
Herring, J.R. 1975 Approach of axisymmetric turbulence to isotropy. {\it Phys. Fluids} {\bf  17}, 859--872.

\bibitem{kerr}
Holm, D.D., and Kerr, R. 2002 Transient vortex events in the initial value problem for turbulence.
{\it Phys. Rev. Lett.} {\bf 88}, 244501.

\bibitem{hopfinger}
Hopfinger, E., Browand, F. \& Gagne, Y. 1982 Turbulence and waves in a rotating tank. 
{\it J. Fluid Mech. } {\bf 125}, 505-534.

\bibitem{iro}
Iroshnikov, P.S.	 1963 Turbulence of a conducting fluid in a strong magnetic field. {\it Sov. Astron.} {\bf 7}, 566-571.

\bibitem{kaneda}
Ishihara, T., Kaneda, Y., Yokokawa, M., Itakura, K. \& Uno, A. 2005
Energy spectrum in the near dissipation range of high solution direct numerical simulation of turbulence.
{\it J. Phys. Soc. Japan} {\bf 74}, 1464--1471.

\bibitem{jacobitz}
Jacobitz, F., Liechtenstein, L., Schneider, K. \& Farge, M. 2008
On the structure and dynamics of sheared and rotating turbulence: 
Direct numerical simulation and wavelet-based coherent vortex extraction.
{\it Phys. Fluids} {\bf 20}, 045103.

\bibitem{jacquin}
Jacquin, L., Leuchter, O., Cambon, C. \& Mathieu, J.  1990 Homogeneous turbulence in the presence of rotation. {\it J. Fluid Mech.} {\bf 220}, 1--52.

\bibitem{K41}
Kolmogorov A.N. 1941
The local structure of turbulence in incompressible viscous fluid for very large Reynolds number. {\it Dokl. Akad. Nauk SSSR}, {\bf 30} 9--13.

\bibitem{RHK65}
Kraichnan, R.H. 1965 Inertial-range spectrum of hydromagnetic turbulence. {\it Phys. Fluids} {\bf  8}, 1385--1387.

\bibitem{kraichnan_1976}
Kraichnan, R.H., 1976
Eddy viscosity in two and three dimensions.
{\it J. Atmos. Sci.}, {\bf 33}, 1521-1536.

\bibitem{kraichnan_montgo}
Kraichnan, R.H. \& Montgomery, D. 1980 Two-dimensional turbulence. {\it Rep. Prog. Phys.} {\bf 43}, 547--619.

\bibitem{KRSTU}
Krstulovic, G., Mininni, P.D., Brachet, M.E. \& Pouquet, A. 2009
Cascades, thermalization and eddy viscosity in helical Galerkin truncated Euler flows.
{\it Phys. Rev. E} {\bf 79}, 056304.

\bibitem{kurien}
Kurien, S., Taylor, M.A., \& Matsumoto, T. 2004
Cascade time scales for energy and helicity in homogeneous isotropic turbulence.
{\it Phys. Rev. E} {\bf 69}, 066313.

\bibitem{lauten88}
Lautenschlager, M., Eppel, D.P. and Thacker, W.C. 1988
Subgrid parametrization in helical flows.
{\it Beitr. Phys. Atmosph.} {\bf 61}, 87-97.

\bibitem{galina}
Levina, G.V. \& Burylov, I.A. 2006
Numerical simulation of helical-vortex effects in Rayleigh-B\'enard convection.
{\it Nonlin. Processes Geophys.} {\bf 13}, 205--222.

\bibitem{lewellen}
Lewellen, D.C. \& Lewellen, W.S. 2007
Near-surface intensification of tornado vortices.
{\it J. Atmos. Sci.} {\bf 64}, 2176-2194.

\bibitem{lilly86}
Lilly, D. 1986 The Structure, Energetics and Propagation of Rotating Convective Storms. Part II: Helicity and Storm Stabilization. {\it J. Atmos. Sci.} {\bf 43}, 126-140.

\bibitem{ma_mazenko}
Ma, S. \& Mazenko, G.F. 1975 
Critical dynamics of ferromagnets in 6-$\epsilon$ dimensions: General discussion and detailed calculation. {\it Phys Rev. B} {\bf 11}, 4077--4100.

\bibitem{markovski}
Markovski, P.M., Straka, J.M.,  Rasmussen, E.N. \& Blanchard, D.O. 1998
Variability of Storm-Relative Helicity during VORTEX.
{\it Monthly weather Rev.}, {\bf 126}, 2959-2971.

  \bibitem{matthaeus}
 Matthaeus, W. H.,  Pouquet, A., Mininni, P. D., Dmitruk, P. \& Breech, B. 2008 Rapid directional alignment of velocity and magnetic field in magnetohydrodynamic turbulence. {\it Phys. Rev. Lett.}{\bf 100}, 085003.
 
\bibitem{McWilliams84}
{Mc{W}illiams, J.~C.} 1984
\newblock The emergence of isolated coherent vortices in turbulent flow.
\newblock {\em J.\ Fluid Mech.} {\bf 146}, 21--43.

\bibitem{pablo_ABC}
 Mininni, P.D., 2007 Inverse cascades and the $\alpha$ effect at a low magnetic Prandtl number.
{\it Phys. Rev. E} {\bf 76}, 026316. 

\bibitem{mininni_rot_hel}
Mininni, P.D. \& Pouquet, A. 2009a
Helicity cascades in rotating turbulence. {\it Phys. Rev. E} {\bf 79}, 026304.

\bibitem{mininni_rot_hel2}
Mininni, P.D. \& Pouquet, A. 2009b
 Rotating helical turbulence. Part I. Global evolution and spectral behavior,
    submitted to {\it Phys. Rev. E},   see also arXiv:0909.1272.

\bibitem{mininni_rot_hel3}
Mininni, P.D. \& Pouquet, A. 2009c
Helical rotating turbulence. Part II. Intermittency, scale invariance and structures,
    submitted to {\it Phys. Rev. E},   see also arXiv:0909.1275.

\bibitem{mininni_rot_TG}
Mininni, P.D., Alexakis, A., \& Pouquet, A. 2009
Scale interactions and scaling laws in rotating flows at moderate Rossby numbers and large Reynolds numbers.
{\it Phys. Fluids} {\bf 21}, 015108.

\bibitem{moffatt}
Moffatt, H.K. 1969 The degree of knottedness of tangled vortex lines. {\it J. Fluid Mech.} {\bf 35}, 117--129.

\bibitem{moffatt2}
Moffatt, H.K. 1983 Transport effects associated with turbulence 
with particular attention to the influence of helicity.
{\it Rep. Prog. Phys.} {\bf 46}, 621--664.

\bibitem{tsinober}
Moffatt, H.K. \& Tsinober A. 1992 Helicity in laminar and turbulent flow. {\it Ann. Rev. of Fluid Mech.} {\bf 24}, 281--312.

\bibitem{TG}
Monchaux, R., Berhanu, M., Bourgoin, M., Moulin, M., Odier, Ph.,  Pinton, J.-F., Volk, R.,  Fauve, S., Mordant, N.,
P\'etr\'elis, F., Chiffaudel, A., Daviaud,  F.,  Dubrulle, B.,  Gasquet, C.,  Mari\'e, L. \& Ravelet,  F. 
2007 Generation of a Magnetic Field by Dynamo Action in a Turbulent Flow of Liquid Sodium.
{\it Phys. Rev. Lett.} {\bf 98}, 044502.

\bibitem{moreau}
Moreau, J.J. 1961 Constantes d'un \^il\^ot tourbillonnaire en fluide parfait barotrope. {\it C. R. Acad. Sci. Paris} {\bf 252}, 2810-- 2812.

\bibitem{morize}
Morize, C.,  Moisy, F. \& Rabaud,  M. 2005 Decaying grid-generated turbulence in a rotating tank.
{\it Phys. Fluids} {\bf 17}, 095105.

\bibitem{Muller07}
M\"uller, W.-C. \& Thiele, M. 2007 Scaling and energy transfer in rotating turbulence . {\it Europhys.\ Lett} {\bf 77} 34003.

\bibitem{naza}
Nazarenko, S. \&  Schekochihin, A. 2009
Critical Balance in Magnetohydrodynamic, Rotating and Stratified Turbulence: Towards a Universal Scaling Conjecture.
Preprint, see also arxiv:0904:3488. 

\bibitem{newell01}
Newell, A.,  Nazarenko, S. \&  Biven, L. 2001 Wave turbulence and intermittency.
{\it Physica D} {\bf 152Ð153}  520Ð550.

\bibitem{emma}  Noether, E. 1918 Invariante Variations probleme. {\it Nachr. d. K\"onig. Gesellsch. d. Wiss. zu G\"ottingen, Math-phys. Klasse}, 235-257; English translation in:  Travel, M.A. 1971{\it Transport Theory and Statistical Physics} {\bf 1}, 183--207.   
 
 \bibitem{olla}
 Olla, P., 1998
 Three applications of scaling to inhomogeneous, anisotropic turbulence.
{\it Phys. Rev. E} {\bf 57}, 2824--2831.

\bibitem{pedlo}
Pedlosky, J. 1986 {\it Geophysical fluid dynamics}, Springer.
          
\bibitem{pelz}
Pelz, R., Yakhot, V. \& Orszag, S.A. 1985 Velocity-vorticity patterns in turbulent flow. 
{\it Phys. Rev. Lett.} {\bf 54}, 2505-2508.

\bibitem{pierrehumbert}
Pierrehumbert, R.T., Held, I.M., \& Swanson, K.L. 1994 Spectra of local and nonlocal 
two-dimensional turbulence, {\it Chaos Solitons Fractals} {\bf 4}, 1111--1116.
 

\bibitem{pouquet_RNG}
Pouquet, A.,  Fournier, J.D. \& Sulem, P. L.1978 Is helicity relevant
for large scale steady state three--dimensional turbulence?  {\it J. Phys.
Lettres} (Paris), {\bf 39}, L 199--203.

\bibitem{patterson}
Pouquet, A., \&  Patterson, G.S. 1978 Numerical simulation of helical
magnetohydrodynamic turbulence. {\it J. Fluid Mech.}, {\bf 85}, 305--323.

\bibitem{moin}
Rogers, M.M., and Moin, P. 1987
The structure of the vorticity field in homogeneous turbulent flows. 
{\it J. Fluid Mech.} {\bf 176}, 33--66.

\bibitem{rotunno_84}
Rotunno, R. 1984
An investigation of a three-dimensional asymmetric vortex.
{\it J. Atmos. Sci.} {\bf 41}, 283--298.

\bibitem{cambon_book}
{Sagaut, P. \& Cambon, C.} 2008
Homogeneous turbulence dynamics.
\newblock Cambridge Univ.\ Press, Cambridge.

\bibitem{sanada}
Sanada, T. 1993 Helicity production in the transition to chaotic flows simulated by Navier-Stokes equation.
{\it Phys. Rev. Lett.} {\bf 70}, 3035--3038.

\bibitem{seiwert}
Seiwert, J., Morize, C.  \&  Moisy, F. 2008
On the decrease of intermittency in decaying rotating turbulence.
{\it Phys. Fluids} {\bf 20}, 071702.

\bibitem{shaw}
Shaw, R.A., and Oncley, S. P. 2001
Acceleration intermittency and enhanced collision 
kernels in turbulent clouds.
{\it Atmospheric Research} {\bf 59-60}, 77-87.

\bibitem{SL}
She, Z-S. \& L\'ev\^eque, E. 1994 Universal scaling laws in fully developped turbulence.
{\it Phys. Rev. Lett.} {\bf 72}, 336--339.

\bibitem{Simand00}
{Simand, C., Chill\`a, F., \& Pinton, J.-F.} 2000
\newblock Study of inhomogeneous turbulence in the closed flow between corotating disks.
\newblock {\em Europhys.\ Lett.} {\bf 49}, 336--342.

\bibitem{Smith96}
Smith, L.~M., Chasnov, J. \& Waleffe, F. 1996
 Crossover from Two- to Three-Dimensional Turbulence.
 {\it Phys.\ Rev. Lett. } {\bf 77}, 2467--2470.
 
\bibitem{Smith99}
Smith, L.~M. \& Waleffe, F. 1999
Transfer of energy to two-dimensional large scales in forced,
  rotating three-dimensional turbulence.
 {\it Phys.\ Fluids } {\bf 11}, 1608--1622.

\bibitem{tabeling02}
Tabeling, P. 2002 Two-dimensional turbulence: a physicist approach. {\em Phys. Rep.} {\bf 362}, 1--62.

\bibitem{waleffe_1993}
Waleffe, F. 1993
Inertial transfers in the helical decomposition.
{\it Phys. Fluids}, {\bf A5}, 677-685.

\bibitem{wurman_science_1996}
Wurman, J.,  Straka, J.M. \& Rasmussen, E.N. 1996
Fine-Scale Doppler Radar Observations of Tornadoes.
{\it  Science}, {\bf 272}, 1774-1777.

\bibitem{wurman_2002}
Wurman,  J.  2002
The Multiple-Vortex Structure of a Tornado.
{\it Weather and Forecasting}, {\bf 17}, 473--505.

\bibitem{yahalom}
Yahalom, A. 1995
Helicity Conservation via the Noether Theorem. 
{\it J. Math. Phys.} {\bf 36}, 1324--1327.

\bibitem{yokoi93}
Yokoi, N. \& Yoshizawa, A., 1993
Statistical analysis of the effects of helicity in inhomogeneous turbulence.
{\it Phys. Fluids}, {\bf A5}, 464--477.

\bibitem{falko_book}
Zakharov, V.E., Lvov, V.S., \& Falkovich, G.G. 1992 {\it Kolmogorov spectra of turbulence}. Springer-Verlag, Berlin.

\bibitem{zeman}
Zeman, O. 1994 A note on the spectra and decay of rotating hoqmgeneous turbulence.
{\it Phys. Fluids} {\bf 6}, 3221--3223.

\bibitem{lohse1}
 Zhong, J-Q, Stevens, R.J.A.M., Clercx, H.J.H., Verzicco, R., 
Lohse, D. \& Ahlers, G. 2009
Prandtl-, Rayleigh-, and Rossby-Number Dependence of Heat Transport 
in Turbulent Rotating Rayleigh-B\'enard Convection.
{\it Phys. Rev. Lett.} {\bf 102}, 044502.

\bibitem{Zhou95}
{Zhou, Y.} 1995
\newblock A phenomenological treatment of rotating turbulence.
\newblock {\em Phys.\ Fluids} {\bf 7}, 2092--2094.

\bibitem{zimmerman}
Zimmerman, W.B. 1996 Fluctuations in passive tracer concentration due to mixing by coherent structures in anisotropic, homogeneous, helical turbulence. {\it IChemE Symposium Series} {\bf 140}, 213--221.

\end{thebibliography}
\end{document}